\documentclass[12pt]{article}
\usepackage{epsfig}

\usepackage{a4}
\usepackage{a4wide}

\def\gs{\mathrel{
   \rlap{\raise 0.511ex \hbox{$>$}}{\lower 0.511ex \hbox{$\sim$}}}}
\def\ls{\mathrel{
   \rlap{\raise 0.511ex \hbox{$<$}}{\lower 0.511ex \hbox{$\sim$}}}}
\newcommand{\obb}{0\mbox{$\nu\beta\beta$}}
\newcommand{\onbb}{neutrino-less double beta decay}
\newcommand{\ba}{\begin{array}{c}}
\newcommand{\baz}{\begin{array}{cc}}
\newcommand{\bad}{\begin{array}{ccc}}
\newcommand{\bav}{\begin{array}{cccc}}
\newcommand{\bea}{\begin{equation} \begin{array}{c}}
\newcommand{\eea}{ \end{array} \end{equation}}
\newcommand{\ea}{\end{array}}
\newcommand{\D}{\displaystyle}
\newcommand{\dms}{\mbox{$\Delta m^2_{\odot}$}}
\newcommand{\dma}{\mbox{$\Delta m^2_{\rm A}$}}
\newcommand{\meff}{\mbox{$\left| m_{ee} \right|$}}
\newcommand{\eV}{\mbox{ eV}}

\newcommand{\be}{\begin{eqnarray}}
\newcommand{\ee}{\end{eqnarray}}

\newcommand{\sss}{\sin^2 \theta_{12}}

\begin{document}

\begin{titlepage}
\title{\vspace*{-2.0cm}
\hfill {\small hep--ph/0703135}\\[20mm]
\bf\Large
Getting Information on $|U_{e3}|^2$ from Neutrino-less Double Beta Decay
\\[5mm]\ }

\author{
Alexander Merle\thanks{email: \tt alexander$.$merle@mpi-hd.mpg.de}~~~and~
Werner Rodejohann\thanks{email: \tt werner$.$rodejohann@mpi-hd.mpg.de} 
\\ \\
{\normalsize \it Max-Planck-Institut f\"ur Kernphysik,}\\
{\normalsize \it Postfach 10 39 80, D-69029 Heidelberg, Germany}
}
\date{}
\maketitle
\thispagestyle{empty}

\begin{abstract}
\noindent
We consider the possibility to gain information on the 
lepton mixing matrix element $|U_{e3}|$ from an improved experimental 
limit on the effective neutrino mass governing neutrino-less double 
beta decay. We show that typically a lower limit on $|U_{e3}|$ as a 
function of the smallest neutrino mass can be set. 
Furthermore, we give the values of the sum of neutrino masses 
and $|U_{e3}|$ which are allowed and forbidden by an experimental upper 
limit on the effective mass. Alternative explanations for neutrino-less 
double beta decay, Dirac neutrinos or unexplained cosmological features 
would be required if future measurements showed that the values lie in 
the respective regions. 
Moreover, we show that a measurement of $|U_{e3}|$ from neutrino-less 
double beta decay is very difficult due to the expected errors on 
the effective mass and the oscillation parameters.

\end{abstract}

\end{titlepage}

\section{\label{sec:intro}Introduction}

In the next decade, neutrino physics \cite{reviews} is bound to answer 
several crucial questions. 
Most interesting are in particular the ones about the magnitude 
of the last unknown mixing parameter $U_{e3}$ and the one about 
the possible Majorana nature of the neutrino. 
The magnitude of $|U_{e3}|$ is crucial for 
the future experimental program of 
neutrino physics, since in neutrino oscillation 
experiments the search for $CP$ violation and partly 
the determination of the neutrino mass hierarchy depends on 
it \cite{fut_osc}. Its value is also an important 
criterion in order to rule out the many models for neutrino masses 
and mixing \cite{albright}. Information on $|U_{e3}|$ can be obtained by 
\begin{itemize}
\item direct searches in reactor neutrino experiments \cite{ue3_reactor}; 
\item studies with future long-baseline neutrino experiments \cite{ue3_LBL}, 
even including $\beta$-beams or neutrino factories 
\cite{ue3_beta};  
\item observation of supernova neutrinos \cite{ue3_SN};  
\item observation of fluxes (or flux ratios) of high-energy neutrinos in
neutrino telescopes \cite{ue3_UHE}. 
\end{itemize}
In what regards the Majorana nature of the neutrino, only 
experiments looking for neutrino-less double beta decay (\obb) \cite{APS} 
are realistic possibilities, 
as other lepton number violating processes are far beyond reach \cite{ich}. 
The life-time of 
(or a lower limit on) \obb~is translated into a value (or an upper limit) 
of the effective mass 
\be
\meff = \left| \sum U_{ei}^2 \, m_i \right|~.
\ee
Here $m_i$ with $i=1,2,3$ are the neutrino masses and 
$U$ is the Pontecorvo-Maki-Nagakawa-Sakata (PMNS) lepton mixing matrix. 
From the Heidelberg-Moscow experiment the current limit on the half-live 
of $^{76}$Ge is $1.9 \cdot 10^{25}$ y at 90 $\%$ C.L.~\cite{KK}, 
translating into a limit on the effective mass 
of $\meff \le 0.35 \, \zeta$ eV, 
where $\zeta$ is of ${\cal O}(1)$ and denotes the nuclear matrix element 
uncertainty. Being the best limit for many years, the values are 
beginning to be challenged by the CUORICINO experiment working 
with $^{130}$Te (T$_{1/2} \ge 3 \cdot 10^{24}$ y) \cite{cuore} 
and the NEMO 3 experiment working with 
$^{100}$Mo (T$_{1/2} \ge 5.8 \cdot 10^{23}$ y) and 
$^{82}$Se (T$_{1/2} \ge 1.2 \cdot 10^{23}$ y) \cite{nemo}. Other approaches such 
as EXO-200 \cite{EXO} or GERDA \cite{Abt:2004yk} will soon do the same.  
Apart from 
proving the Majorana nature 
of neutrinos, the value of \meff~depends on 7 of the 9 parameters 
of low energy neutrino physics, which implies that a large 
amount of information is encoded in a measurement of or a limit on the 
effective mass. Furthermore, in the charged 
lepton basis, \meff~is the $ee$ element of the neutrino mass matrix. 
These interesting 
features render the effective mass  -- in connection 
with future precision measurements 
of the other parameters -- also a very helpful criterion to rule out 
models \cite{Choubey:2005rq}. 

\begin{figure}[ht]
\begin{center}
\epsfig{file=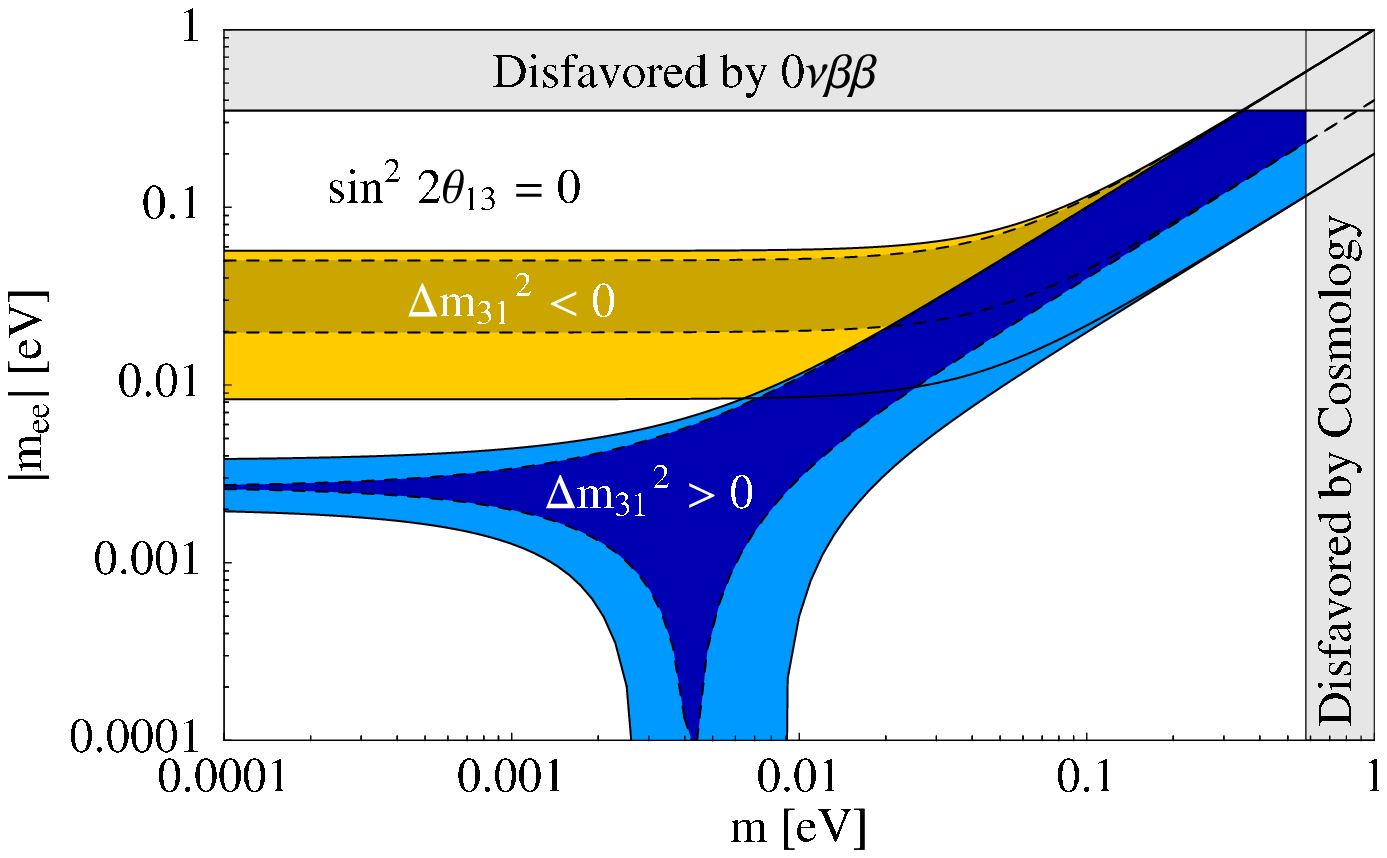,width=8cm,height=5.6cm}
\epsfig{file=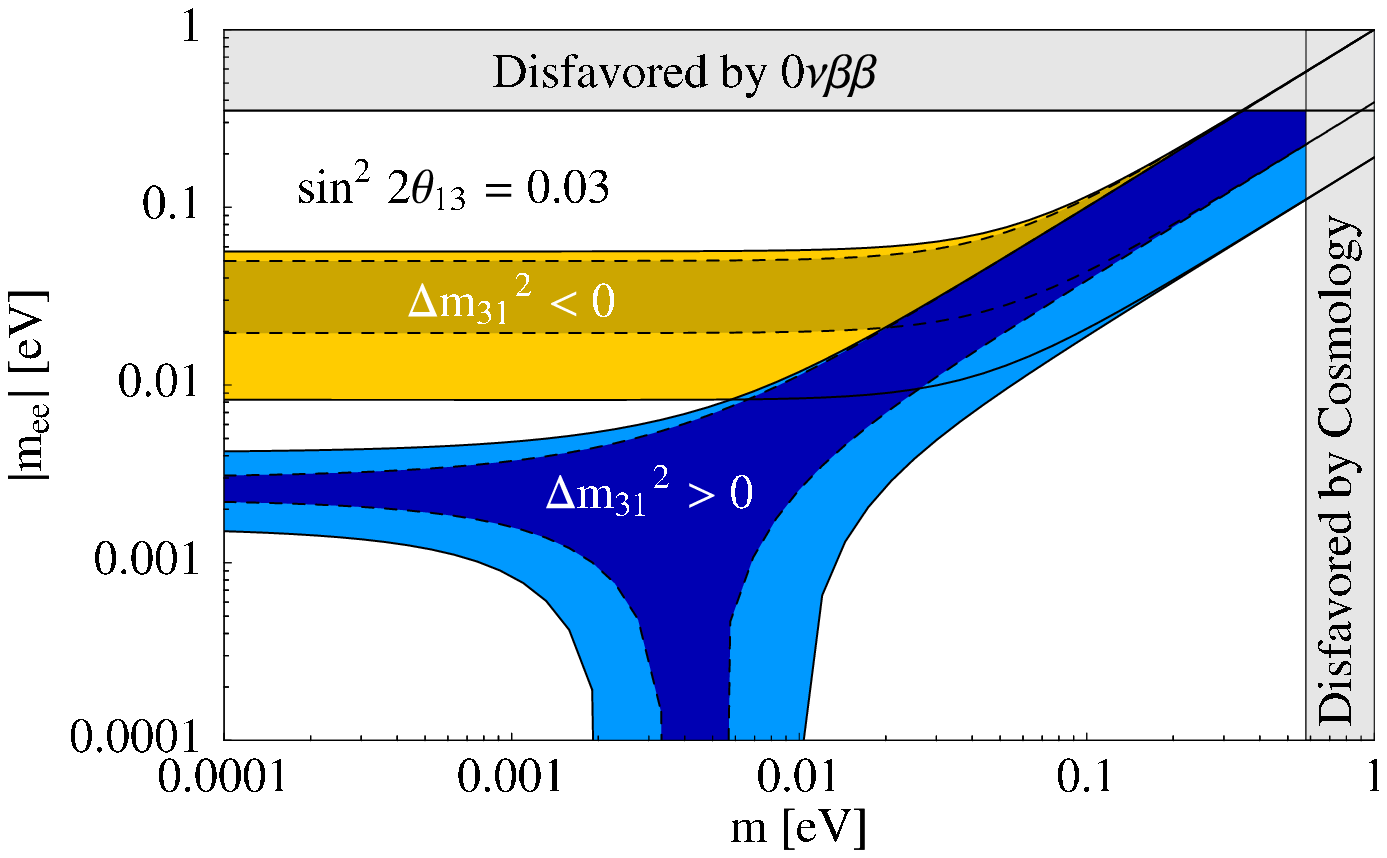,width=8cm,height=5.6cm}
\epsfig{file=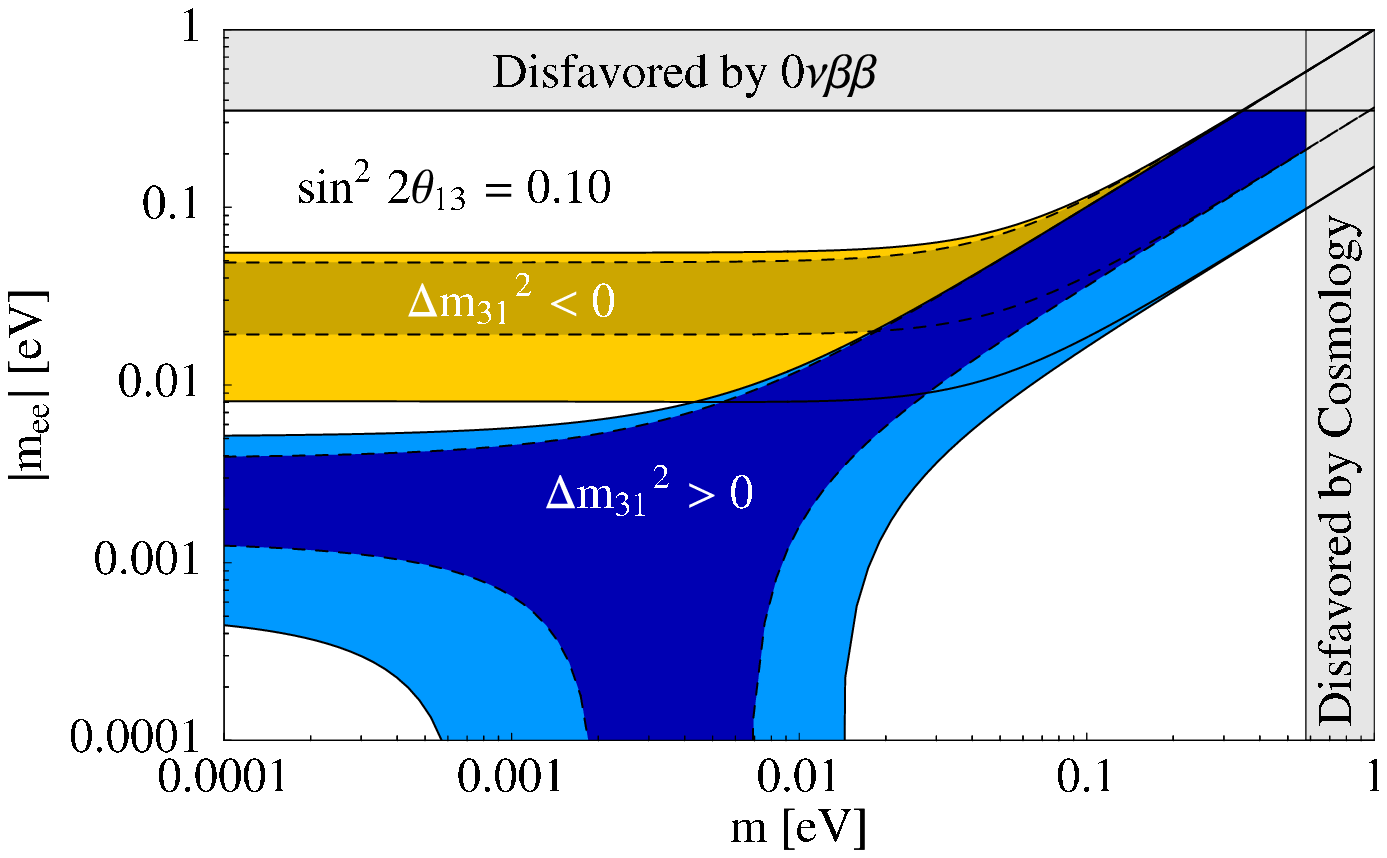,width=8cm,height=5.6cm}
\epsfig{file=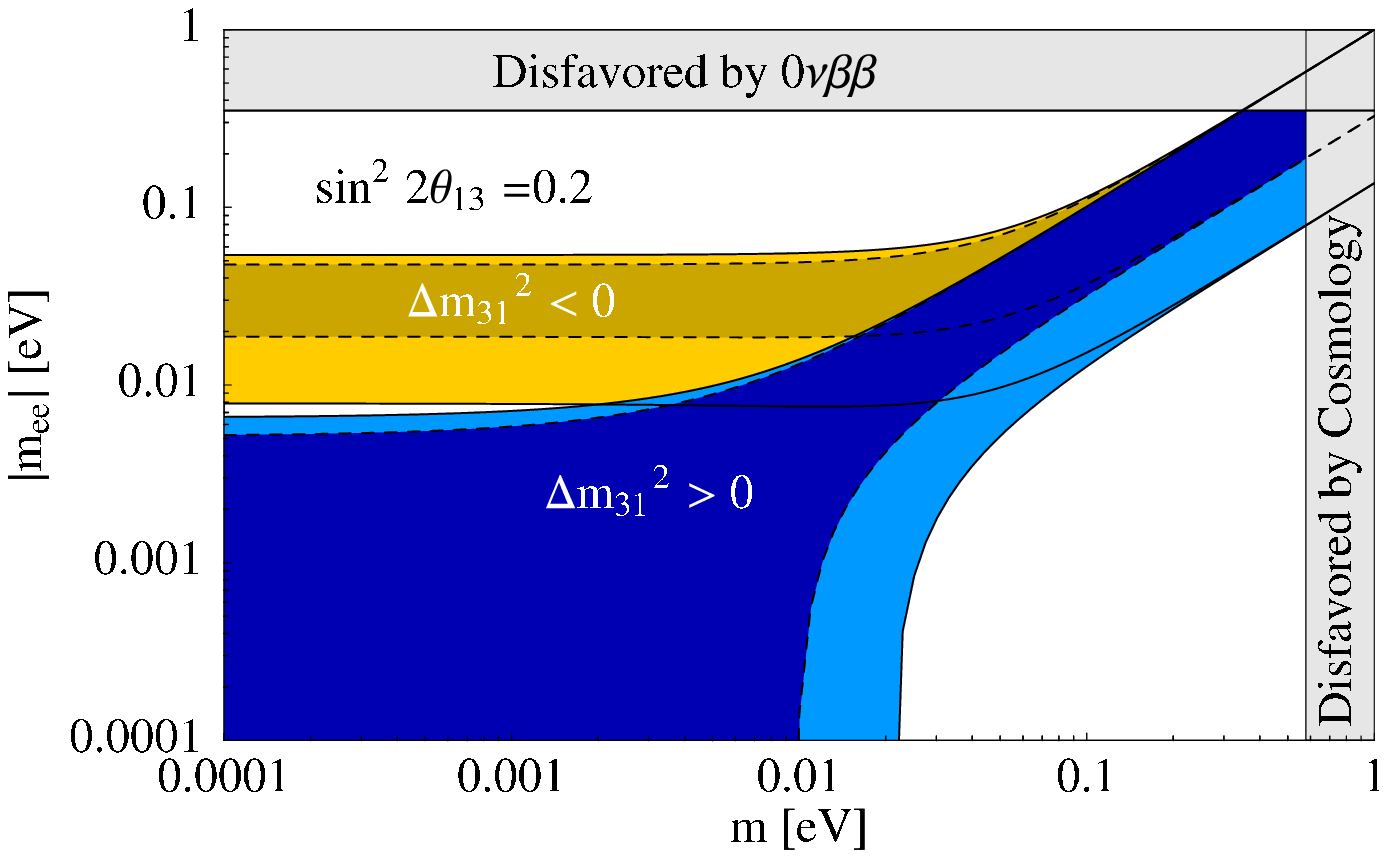,width=8cm,height=5.6cm}
\caption{\label{fig:meff}The effective mass as a function of the 
smallest neutrino mass for the normal and inverted ordering 
and for different values of $|U_{e3}|^2$. 
The dark regions are for the best-fit values of 
the other oscillation parameters, the light regions for their $3\sigma$ 
values.}
\end{center}
\end{figure}

Many authors have studied the predictions for 
\meff~\cite{Choubey:2005rq,eff,Minakata:2001kj,PPR0,Lindner:2005kr,Dev:2006if}
and how a measurement or an improved limit of 
\onbb~will allow to probe the parameters of neutrino 
physics, in particular the 
scale and ordering of neutrino masses and the $CP$ 
parameters\footnote{The predictions for the 
remaining five elements of the neutrino mass matrix were studied in 
Ref.~\cite{wir}.}. 
A recent review can be found in Ref.~\cite{eff_rev}. 
Here we will concentrate on the 
possibility \cite{Minakata:2001kj} to obtain information 
on the parameter $|U_{e3}|^2$ from an improved limit 
on the effective mass. In case of a vanishing effective mass, this has 
been first discussed in Ref.~\cite{Dev:2006if}. 

To motivate this analysis, we show in 
Fig.~\ref{fig:meff} the usual plot of the effective mass as a 
function of the smallest neutrino mass. Both possible signs of the 
atmospheric mass-squared difference are taken, and the best-fit values 
of $\theta_{12}$, \dms, and \dma\ as well as their currently allowed 
$3\sigma$ ranges are used. We have chosen four different 
values of $|U_{e3}|^2$, and the difference between these cases is 
obvious \cite{Lindner:2005kr}. Therefore the possibility to obtain information on 
$|U_{e3}|^2$ via \onbb~is definitely present. As we will elaborate on 
in the course of this letter, 
a lower limit on $|U_{e3}|^2$ as a function of the smallest neutrino 
mass can be set from an upper limit on the effective mass \meff. 
We partly update the results from Ref.~\cite{Minakata:2001kj}. 
Moreover, regions in the parameter space of 
$|U_{e3}|^2$ and the sum of neutrino masses (as measurable in cosmology), 
which contain their allowed and forbidden values, are identified. 
This could be used, for instance, to perform a 
consistency check of future measurements of neutrino masses, 
$|U_{e3}|^2$, and \meff~in order to rule out or constrain alternative 
mechanisms of \onbb, to show that neutrinos are Dirac particles, or to 
find inconsistencies in cosmological methods and models. 
In Ref.~\cite{Lindner:2005kr} it was shown how the value of 
$|U_{e3}|^2$ influences the possibility to distinguish the normal 
from the inverted mass ordering with a measurement of or limit on 
neutrino-less double beta decay. Here we turn round the argumentation and 
investigate the information on $|U_{e3}|^2$ which can be 
obtained by neutrino-less double beta decay. 
All in all, the dependence of \meff~on $|U_{e3}|^2$ is weaker than 
on the other relevant parameters. It is however a 
worthy exercise to investigate inasmuch 
the two most interesting parameters of neutrino physics, $|U_{e3}|^2$ 
and the effective mass, are phenomenologically connected.  
We finally quantify for the 
first time that a measurement of  $|U_{e3}|^2$ via \obb~is extremely 
challenging due to the expected experimental errors. We should remark here that 
the implications we are discussing rely -- of course -- 
on neutrinos being Majorana particles.\\

To set the stage of our discussion, we parameterize the PMNS matrix as 
\be 
U = \left( \bad 
c_{12} \, c_{13} & s_{12} \, c_{13} & s_{13} \, e^{-i \delta}  \\[0.2cm] 
-s_{12} \, c_{23} - c_{12} \, s_{23} \, s_{13} \, e^{i \delta}
& c_{12} \, c_{23} - s_{12} \, s_{23} \, s_{13} \, e^{i \delta}
& s_{23} \, c_{13}  \\[0.2cm] 
s_{12} \, s_{23} - c_{12} \, c_{23} \, s_{13} \, e^{i \delta} & 
- c_{12} \, s_{23} - s_{12} \, c_{23} \, s_{13} \, e^{i \delta}
& c_{23} \, c_{13}  \\
               \ea   \right)
 {\rm diag}(1, e^{i \alpha}, e^{i (\beta + \delta)})~, 
\label{eq:Upara}
\ee 
where $s_{ij} = \sin \theta_{ij}$, $c_{ij} = \cos \theta_{ij}$. 
We have one Dirac phase $\delta$ and two Majorana phases \cite{MajPha} 
$\alpha$ and $\beta$. 
The following best-fit values and $3\sigma$ ranges of the 
oscillation parameters have been obtained \cite{thomas}: 
\begin{eqnarray}
\dms = m_2^2 - m_1^2 &=& 
\left(7.9^{+1.0}_{-0.8}\right) 
\cdot 10^{-5} \eV^2~,\nonumber\\
\sss &=& 0.30^{+0.10}_{-0.06} ~,\nonumber\\
\dma = \left| m_3^2 - m_1^2 \right|&=&  
\left(2.5^{+0.7}_{-0.6}\right) 
\cdot 10^{-3} \eV^2~,\\
\sin^2\theta_{23} &=& 0.50^{+0.18}_{-0.16} ~,\nonumber\\
|U_{e3}|^2  &=&0^{+0.041}_{-0.000} ~.\nonumber
\end{eqnarray}
Crucial for the form of the neutrino mass matrix and the 
value of the effective mass is the 
mass ordering, which can be normal or inverted:
\be
\label{eq:masses}
\bav
\mbox{normal:}  & m_3 > m_2 > m_1 & \mbox{with} & m_2 = \sqrt{m_1^{2}+\dms} ~;
~~~~~~  m_3 = \sqrt{m_1^{2}+\dma} ~,\\[0.3cm]
\mbox{inverted:} & m_2 > m_1 > m_3 &  \mbox{with} & 
m_2 = \sqrt{m_3^{2}+\dms+\dma} ~;~~~~ 
m_1 = \sqrt{m_3^{2} + \dma} ~.
\ea
\ee
Of special interest are the following three extreme cases:  
\be \label{eq:nh}
\mbox{ normal hierarchy~(NH):} 
& ~~~~~~~~~~~~~ 
m_3 \simeq \sqrt{\dma} \gg m_{2} \simeq \sqrt{\dms} \gg m_1~,\\[0.3cm]
\mbox{ inverted hierarchy~(IH):} \label{eq:ih}
& m_2 \simeq m_1 \simeq \sqrt{\dma} \gg m_{3} ~,\\[0.3cm]
\mbox{ quasi-degeneracy~(QD):} \label{eq:qd}
& ~~~~~~~~ m_0^2 \equiv m_1^2 \simeq m_2^2 \simeq m_3^2  \gg \dma~.
\label{eq:mass}
\ee
Having defined the framework, we turn now to the possibility to gain 
information on $|U_{e3}|^2$ from neutrino-less double beta decay.

\section{\label{sec:LowLim}Constraints on $|U_{e3}|^2$ from 
Neutrino-less Double Beta Decay}
Within our parametrization from Eq.~(\ref{eq:Upara}), the 
effective mass as measurable in \onbb\ is given by
\be
\meff = \left|m_1 \, c_{12}^2 \, c_{13}^2 + 
m_2 \, s_{12}^2 \, c_{13}^2 \, e^{2i \alpha}+ m_3 \, s_{13}^2 \, 
e^{2i \beta}\right|~.
\label{eq:effmass}
\ee
With writing Eq.~(\ref{eq:effmass}) as 
\bea
\meff = c_{13}^2 \, 
\left| f(m_1, m_2, \theta_{12}, \alpha, \beta) 
+ m_3 \, \tan^2 \theta_{13} \right| \\[0.2cm]\mbox{ with } 
f(m_1, m_2, \theta_{12}, \alpha, \beta) 
= \left( 
m_1 \, c_{12}^2 + m_2 \, s_{12}^2 \, e^{2i\alpha} 
\right) \, e^{-2 i \beta}~,
\label{eq:effmassmult}
\eea
it is obvious that from an experimental upper limit on \meff\ a lower 
limit on $|U_{e3}|^2$ can be obtained if 
\cite{Minakata:2001kj}
\be \label{eq:defA}
|f(m_1, m_2, \theta_{12}, \alpha, \beta)| \geq m_3 \, \tan^2 \theta_{13}~.
\ee
Note that the absolute value of the function $f$, 
and therefore the lower limit on $|U_{e3}|^2$, does not depend on the 
Majorana phase $\beta$. We now show that for the inverted mass ordering and for 
quasi-degenerate neutrinos the condition (\ref{eq:defA}) 
is always fulfilled: using that $\dms = m_2^2 - m_1^2$ and 
$\dma = |m_3^2 - m_1^2|$, it is easy to see that 
\be \label{eq:fIHQD}
|f| \ge m_1 - s_{12}^2 \, (m_2 + m_1) \simeq \left\{
\baz
\sqrt{\dma} \, \left(\cos 2 \theta_{12} \, (1 - \frac 12 \, \eta) + 
\frac 12 \, s_{12}^2 \, R \right) & \mbox{for IH}~, \\[0.2cm]
m_0 \, \left(c_{12}^2 - s_{12}^2 \, (1 + r_\odot) \right) & \mbox{for QD}~.
\ea 
\right.
\ee
We have defined here the small parameters $R \equiv \dms/\dma$,  
$\eta \equiv m_3^2/\dma$ (in case of IH), and 
$r_\odot = \frac 12 \, \dms/m_0^2$ (in case of QD). 
The left-hand side of Eq.~(\ref{eq:defA}) reads for IH 
$\sqrt{\dma} \, \sqrt{\eta} \, \tan^2 \theta_{13}$ and for QD 
$m_0 \, (1 \pm r_{\rm A}) \, \tan^2 \theta_{13}$, where 
$r_{\rm A} \equiv \frac 12 \, \dma/m_0^2$, the plus (minus) sign is for 
a normal (inverted) ordering, and we have chosen $m_1 = m_0$. 
These values are much smaller than the corresponding values 
in Eq.~(\ref{eq:fIHQD}). 
Therefore the condition in Eq.~(\ref{eq:defA}) is always fulfilled 
for IH and QD, and a lower limit on $|U_{e3}|^2$ can always be obtained 
in these cases.

\begin{figure}[t]
\begin{center}
\epsfig{file=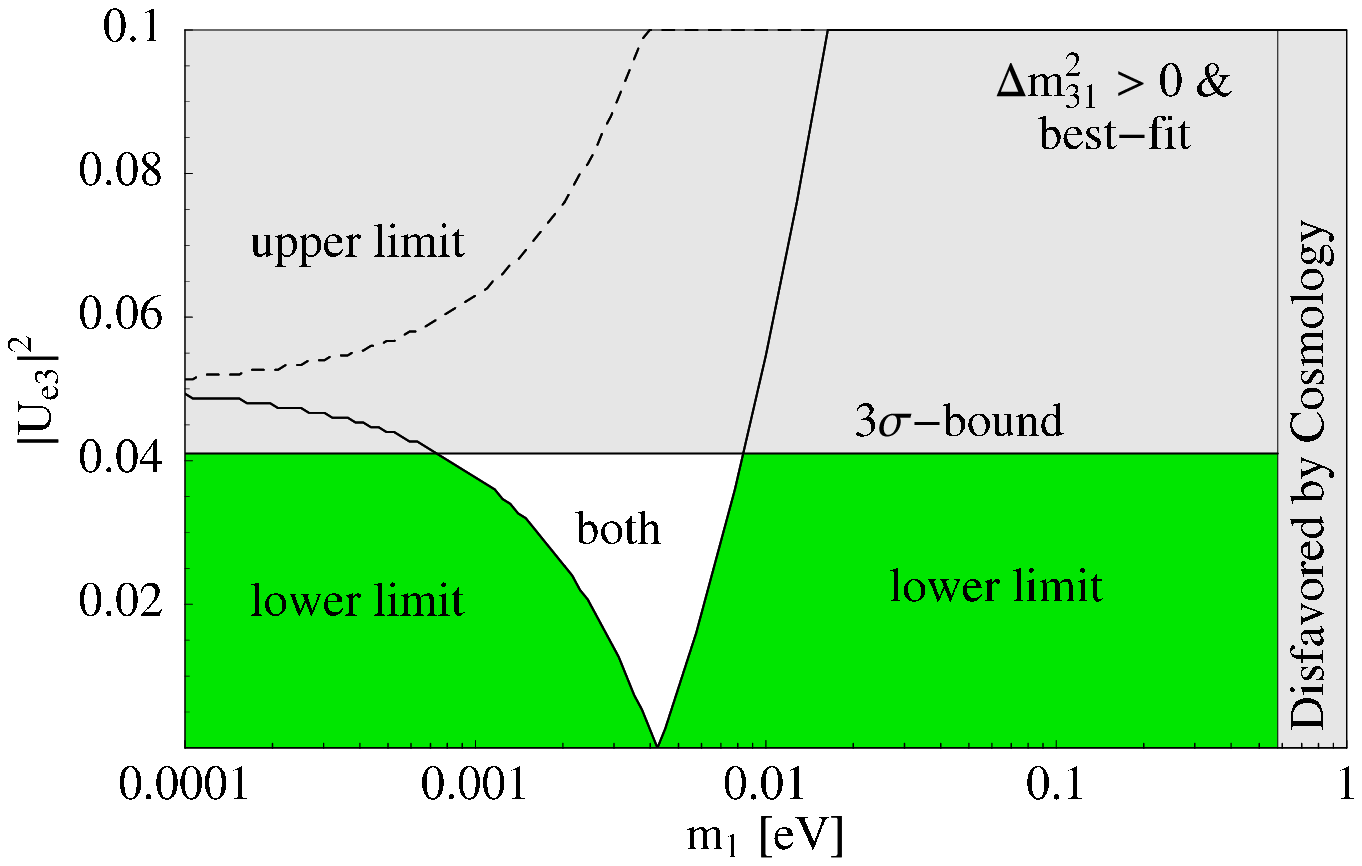,width=8cm,height=6cm}
\epsfig{file=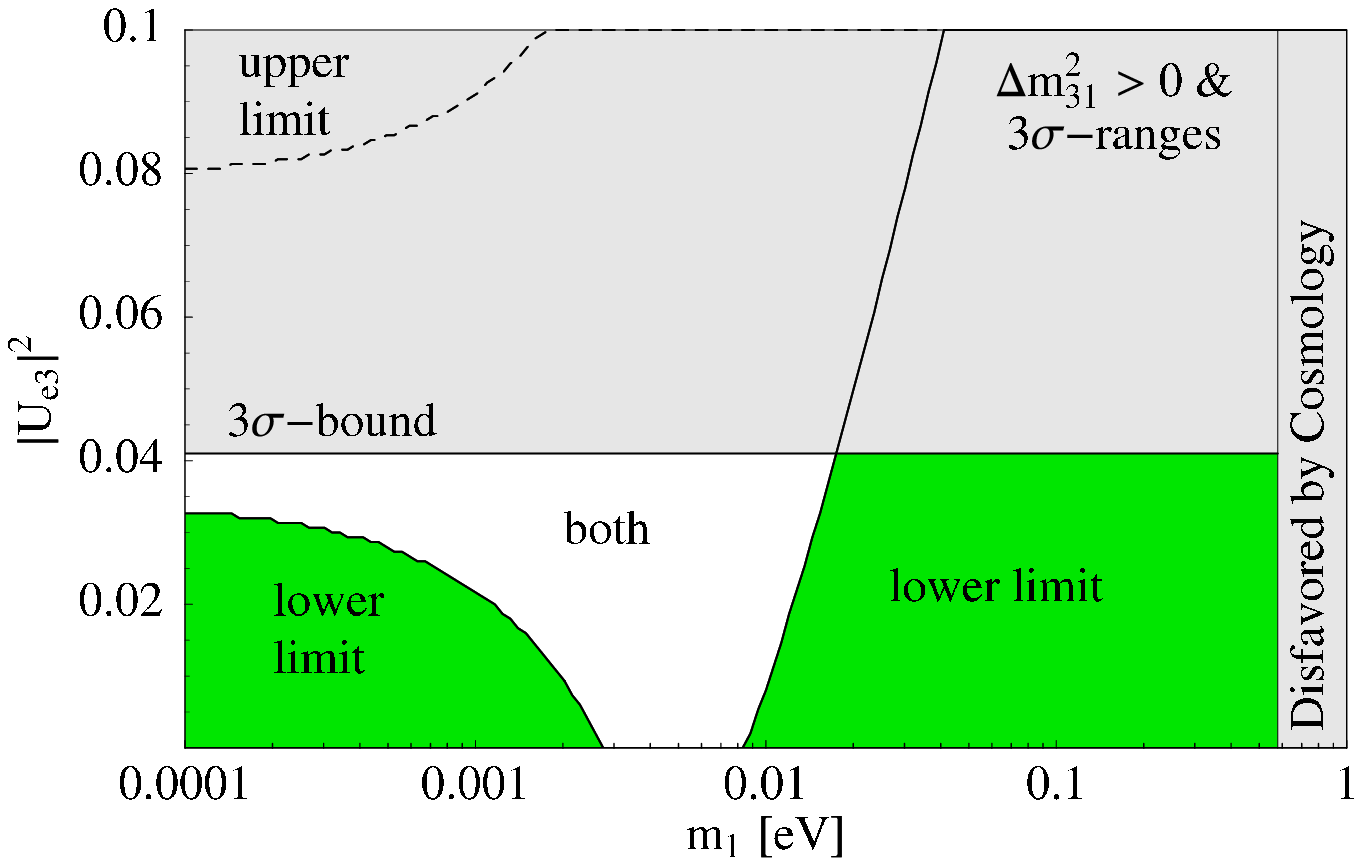,width=8cm,height=6cm}
\caption{\label{fig:allowed}Regions in the $m_1$-$|U_{e3}|^2$-plane 
in which upper and lower limits on $|U_{e3}|^2$ can be obtained. In the 
white region both is possible, depending in particular on the Majorana 
phases. The left plot is for the best-fit values of the oscillation parameters, 
the right plot for the current $3\sigma$ ranges, in both we have also 
indicated the current $3\sigma$ bound on $|U_{e3}|^2$.}
\end{center}
\end{figure}

The situation is different in case of a normal hierarchy. For instance, 
if we neglect $m_1$, then Eq.~(\ref{eq:defA}) translates into 
$\sqrt{\dma} \, \tan^2 \theta_{13} \le \sqrt{\dms} \, \sin^2 \theta_{12}$. 
The $3\sigma$-limit on $\tan^2 \theta_{13}$ is 0.043. 
For the best-fit values of $\theta_{12}$ and the mass-squared differences 
the condition (\ref{eq:defA}) is fulfilled. Consequently, a lower limit on 
$|U_{e3}|^2$ could be set. However, for the values 
$\dms = 7.4 \cdot 10^{-5}$ eV$^2$, $\dma = 2.9 \cdot 10^{-3}$ eV$^2$, and  
$\sin^2 \theta_{12} = 0.25$, the condition (\ref{eq:defA}) 
is not fulfilled, i.e., an upper limit on $|U_{e3}|^2$ can be 
obtained \cite{Minakata:2001kj}.   
The lesson is that there are situations in which small offsets of the 
values of the oscillation parameters lead to different scenarios in what 
regards a limit on $|U_{e3}|^2$. This is of course also true if 
$m_1$ is small but non-zero, in which case the values of both Majorana 
phases also play a crucial role. In Fig.~\ref{fig:allowed} we show in green 
(or, in a black and white printout, dark grey) the regions 
in the $m_1$-$|U_{e3}|^2$-plane in 
which a lower limit on $|U_{e3}|^2$ can be obtained. In the white  
region both, a lower and an upper limit are possible, depending 
on the precise values of $m_{1,2,3}$, $\alpha$, $\beta$, and 
$\theta_{12}$. If $m_1$ 
approaches values in which neutrinos become quasi-degenerate, then 
-- as discussed above -- a lower limit is always obtainable. 
The left plot is for the best-fit 
values of $\sin^2 \theta_{12}$ and the mass-squared differences, the 
right plot for their currently allowed $3\sigma$-ranges. 
Note that there are also regions in which regardless of the values of 
$m_1$, $\alpha$, and $\theta_{12}$ an upper limit can be obtained. 
They are located in the upper left corner of the plots, above the 
dashed line. We have calculated this upper limit for various possible 
bounds on \meff. As it turns out, even for an extremely small bound 
of $1\cdot 10^{-4}$~eV the upper limit lies well above the current 
experimental limit on $|U_{e3}|^2$. Therefore, this possibility will 
not be considered.

For the remainder of the paper we will always assume that the condition 
Eq.~(\ref{eq:defA}) is fulfilled so that a lower limit 
 on $|U_{e3}|^2$ can result. One should keep in mind 
that for part of the $m_1$-$|U_{e3}|^2$-plane in case of a  
normal hierarchy (i.e., small $m_1$), this condition might not be fulfilled. 
However, such small values of $m_1$ will be measured not 
before the very far future \cite{APS}.\\

As clear from the above discussion, and in particular from 
Eq.~(\ref{eq:defA}), an experimental bound on the effective mass leads to
\be
\meff\geq c_{13}^2\, |f|-m_3\, s_{13}^2~.
\label{eq:meelimit}
\ee
This translates into a lower bound on $|U_{e3}|^2$, reading 
\be
|U_{e3}|^2 \ge \frac{|f| - \meff}{|f| + m_3}~, 
\label{eq:lowerlimit}
\ee
where $|f|$ is explicitly given by 
\be \label{eq:deff}
|f|^2 = 
m_1^2 \, c_{12}^4 + m_2^2 \, s_{12}^4 + 
2 \cos 2\alpha \, m_1 \, m_2 \, s_{12}^2 \, c_{12}^2~.
\ee
The lowest allowed $|U_{e3}|^2$ is independent on the Majorana phase 
$\beta$, but a function of the Majorana phase 
$\alpha$. We are interested here in the parameter space of 
neutrino mass and $|U_{e3}|^2$. 
It turns out that the covered area in this parameter 
space is maximized for $\alpha = 0$, and minimized for $\alpha = \pi/2$. 
As the Majorana phase is unknown, one has to take the weakest condition, 
i.e., $\alpha = \pi/2$. This is a special value, as in case of an inverted hierarchy it 
is associated with the conservation of the flavor symmetry 
$L_e - L_\mu - L_\tau$ \cite{lelmlt} and for IH and QD it guarantees stability 
of $\theta_{12}$ with respect to radiative corrections \cite{RGalpi2}. 
Moreover, large cancellations in the effective mass are resulting:
\[ 
\meff_{\alpha=\pi/2} \simeq \left\{ 
\baz
\sqrt{\dma} \, c_{13}^2 \, \cos 2 \theta_{12} & \mbox{ IH~, } \\[0.2cm]
m_0 \, 
\left( 
\cos 2 \theta_{12} + |U_{e3}|^2 \left(\cos 2 \beta - \cos 2 \theta_{12} 
\right) \right) & \mbox{ QD~, } 
\ea
\right. 
\]
to be compared with the maximal possible values 
$ \sqrt{\dma} \, c_{13}^2 $ and $m_0$, respectively.
 
\begin{figure}[tb]
\begin{tabular}[h]{lr}
\epsfig{file=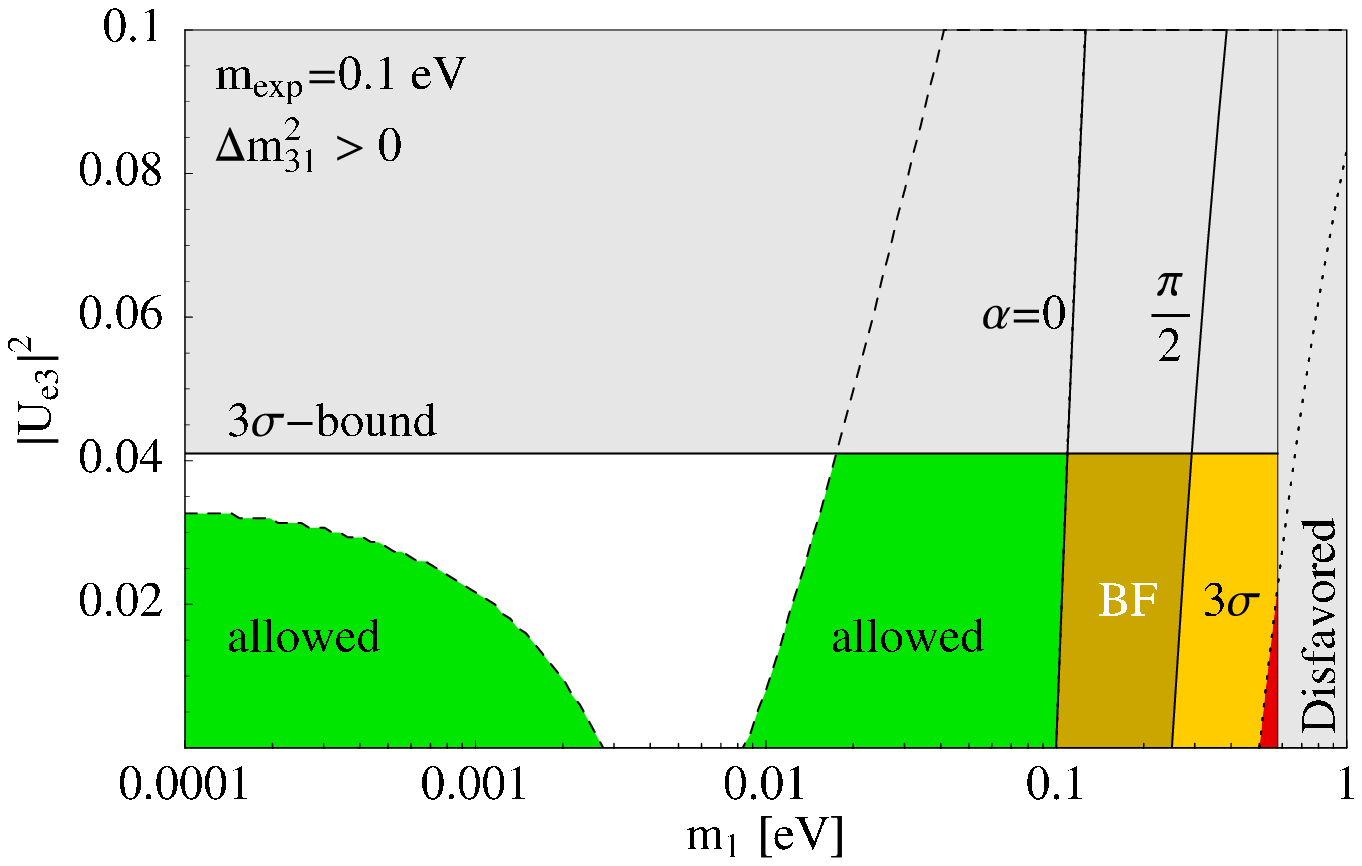,width=8cm}  &
\epsfig{file=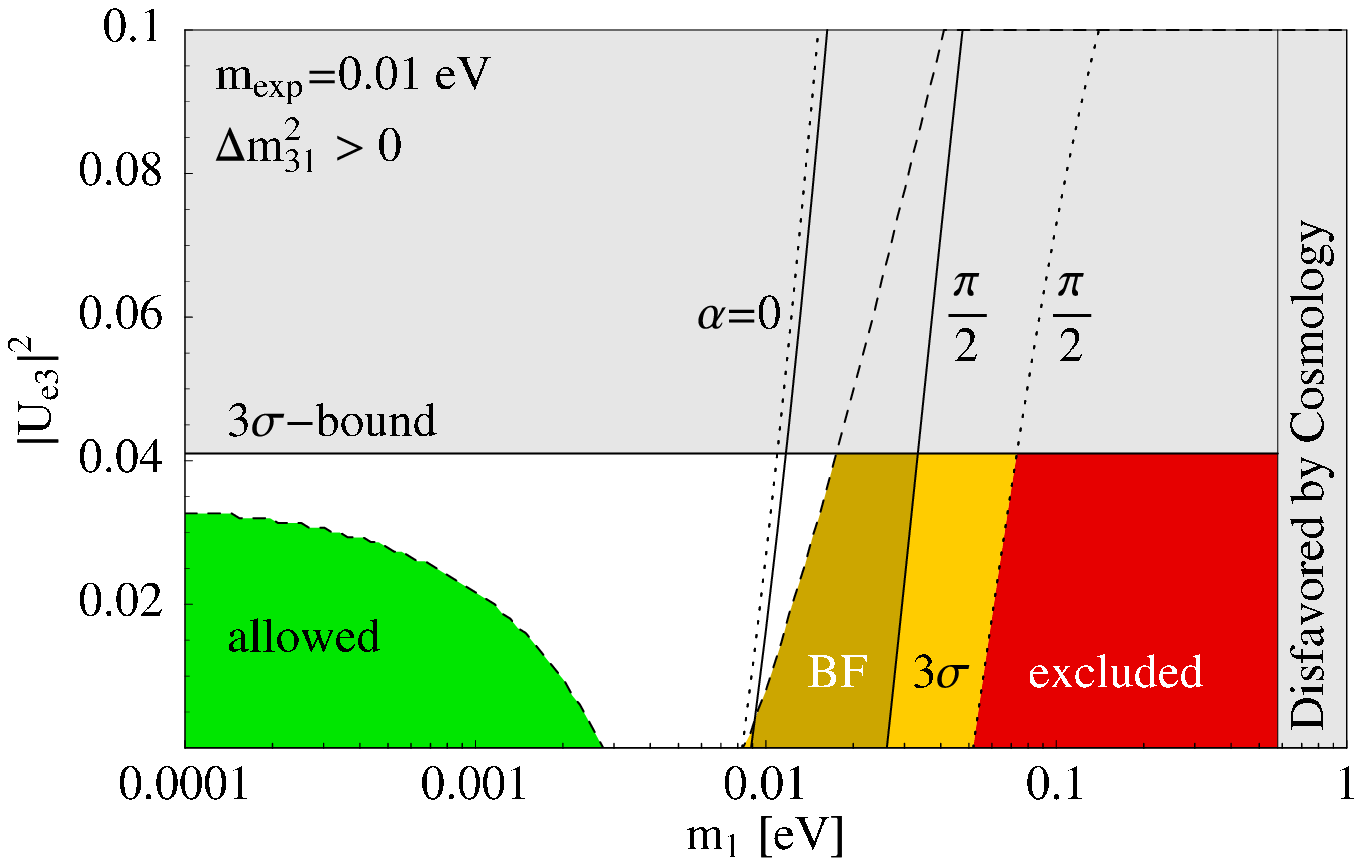,width=8cm} \\
\epsfig{file=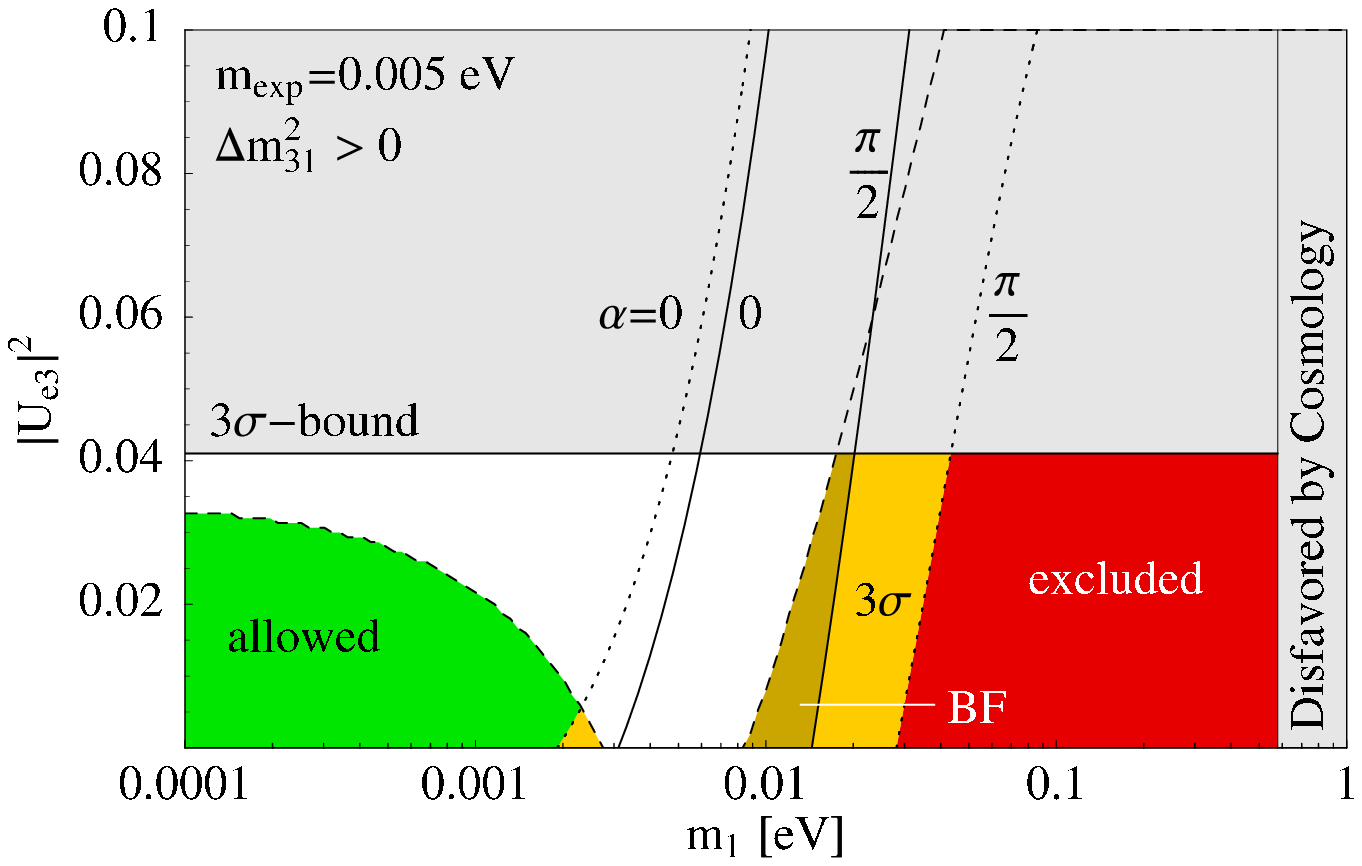,width=8cm} &
\epsfig{file=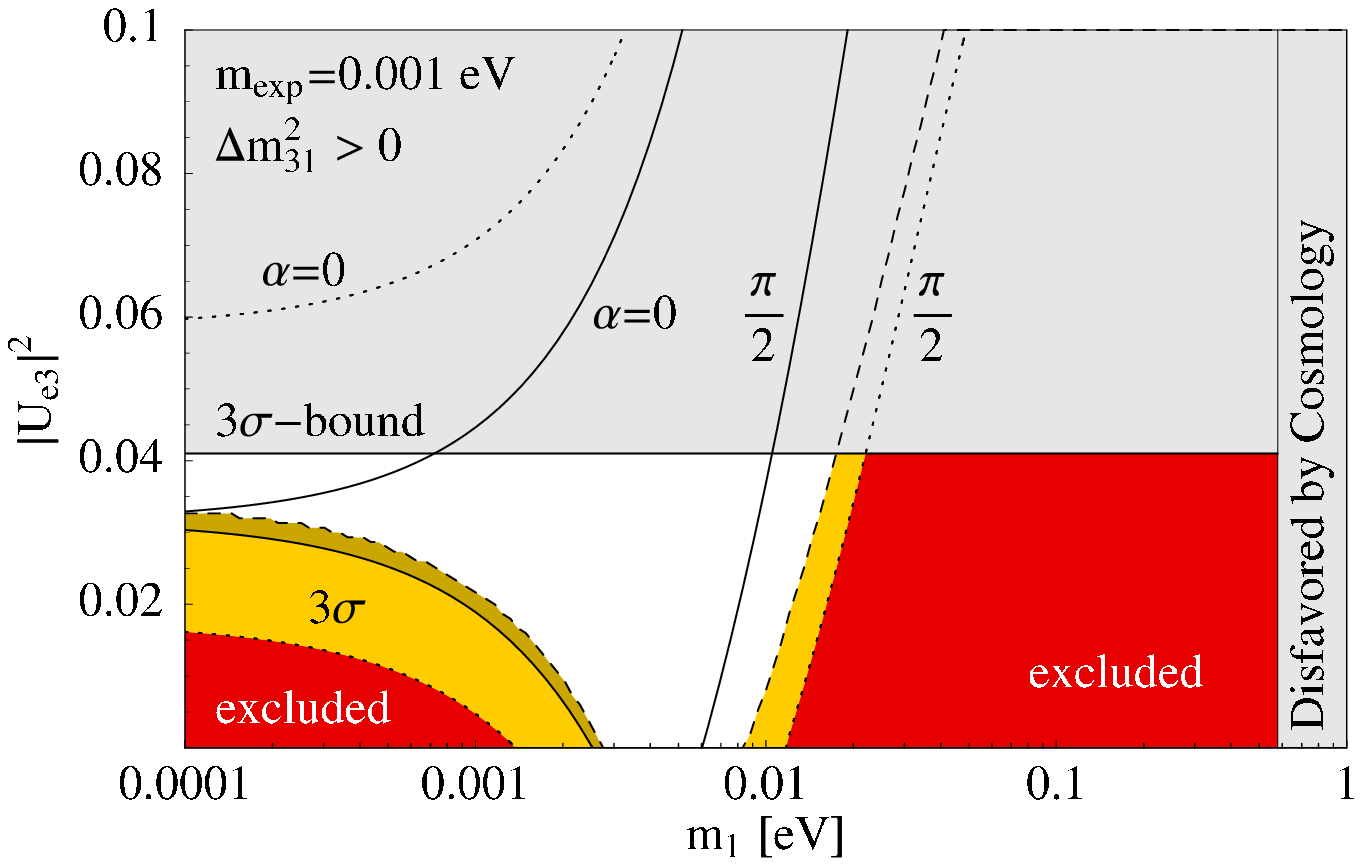,width=8cm} \\
\epsfig{file=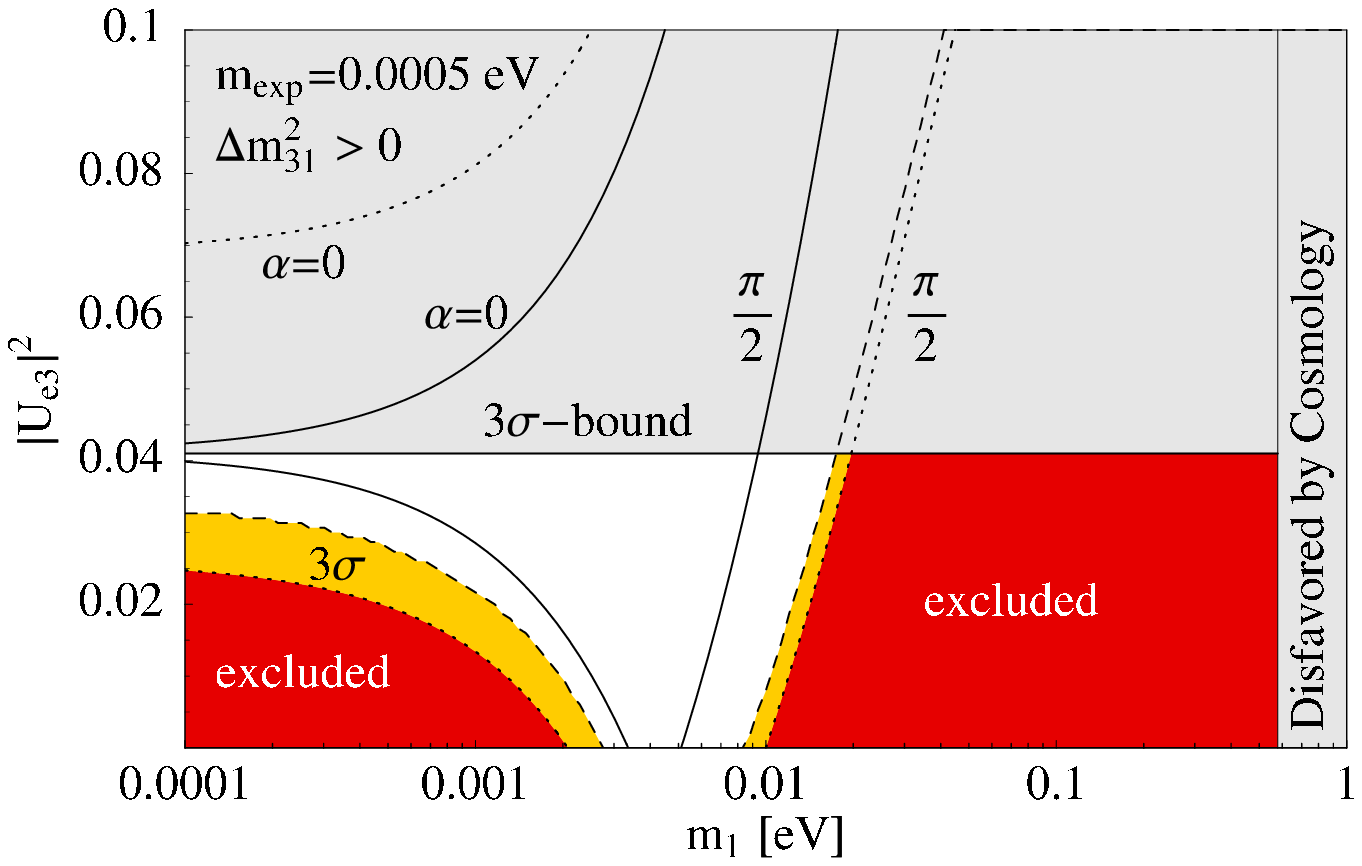,width=8cm} &
\epsfig{file=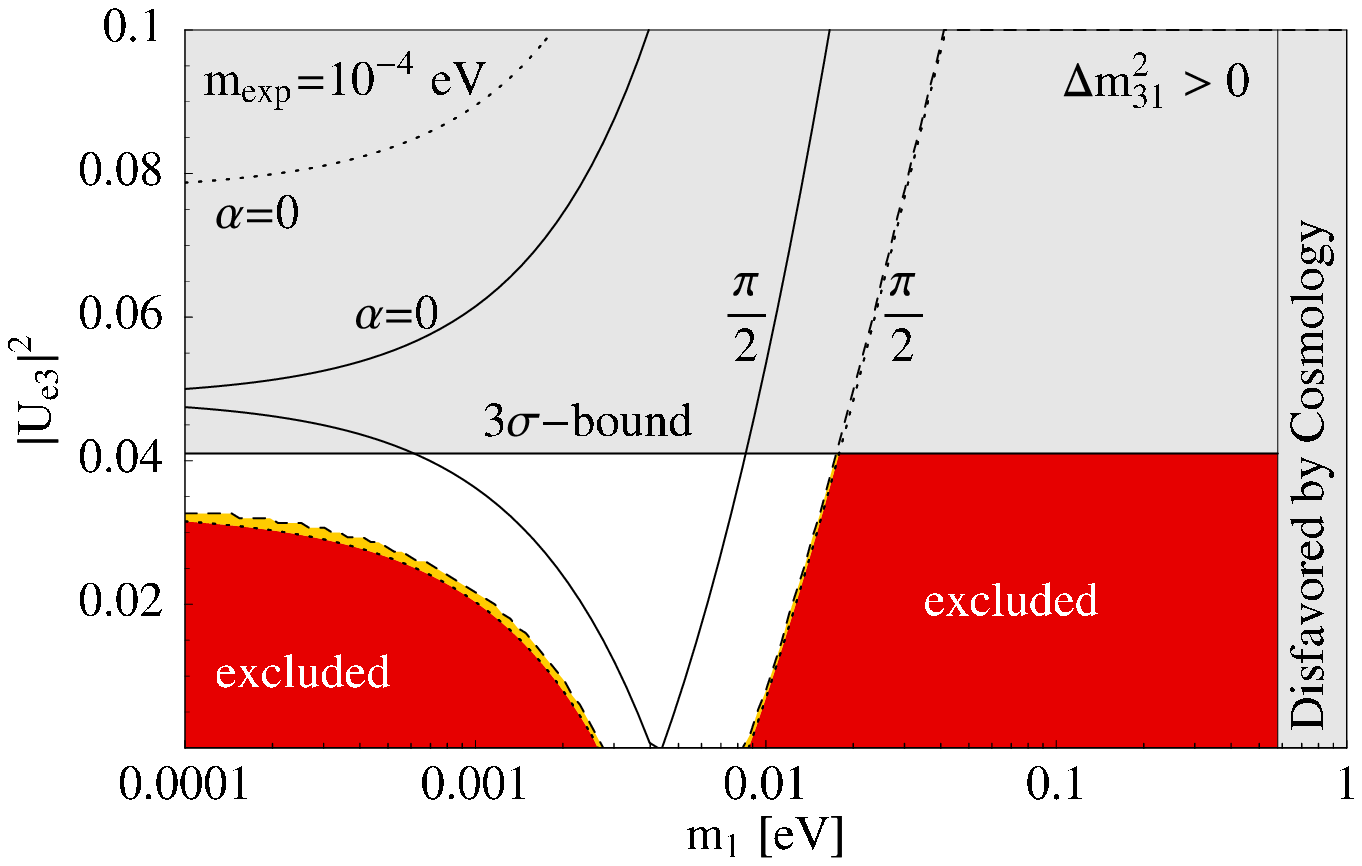,width=8cm}
\end{tabular}
\caption{\label{fig:NHlimits}Normal neutrino mass ordering: 
Areas in the $m_1$-$|U_{e3}|^2$ parameter space in which information 
from \onbb~experiments could be gained. 
Different experimental bounds $m_{\rm exp}$ on the effective mass are shown, 
as well as the current $3\sigma$ limit on $|U_{e3}|^2$. 
The range of the lower limit on $|U_{e3}|^2$ is shown within solid
lines in dark yellow (darkest grey) for the best-fit values (``BF'') 
of the oscillation parameters. 
The light yellow (lightest grey) area within the dotted lines is the 
lower limit for the $3\sigma$-ranges. 
The left and right borders of these areas correspond to $\alpha = 0$ 
and $\alpha = \pi/2$, respectively. 
The red (darkest) areas are excluded, the green (medium grey) and white 
areas are allowed.}
\end{figure}

It is instructive to evaluate the bound 
(\ref{eq:lowerlimit}) in case of the three extreme mass 
hierarchies NH, IH and QD: 
\begin{itemize}
\item[\bf NH:] 
using $m_3^2 \simeq \dma \, (1 + \eta)$ and 
$m_2^2 \simeq \dma \, (R + \eta)$, where 
$\eta = m_1^2/\dma$ and $R = \dms/\dma$, it follows 
\be
\hspace{-.6cm}|U_{e3}|^2 \gs 
\frac{w -\meff/\sqrt{\dma} }{1 + \frac 12 \, \eta + w}~,
\mbox{ where } w = \sqrt{\eta \, c_{12}^4 + (R + \eta) \, s_{12}^4 + 2 \, 
c_{2 \alpha} \, \sqrt{\eta \, R} \, s_{12}^2 \, c_{12}^2}~.
\label{eq:lowerlimitNH}
\ee 
In the extreme case we can neglect $\eta$ and get 
\be
|U_{e3}|^2 \gs \sqrt{R} \, s_{12}^2 - \frac{\meff}{\sqrt{\dma}}~.
\label{eq:NHstrict}
\ee
Recall that for certain values of $\sqrt{R}$ and $s_{12}^2$ a lower limit 
is not possible. 
\item[\bf IH:] for the inverted mass hierarchy 
it holds $m_1^2 = m_3^2 + \dma $ 
and $m_2^2 = m_1^2 + \dms = \dma + \dms + m_3^2$. Neglecting $m_3$ and 
$\dms$ it follows 
\be
|U_{e3}|^2 \gs 
1 - \frac{\meff}{\sqrt{\dma} \, \sqrt{1 - \sin^2 2 \theta_{12} \, 
\sin^2 \alpha} }~.
\label{eq:IHstrict}
\ee
\item [\bf QD:] in the region of quasi-degenerate neutrinos, using 
$m_0 \equiv m_1 \simeq m_2 \simeq m_3$, we have 
\be
|U_{e3}|^2 \gs 
\frac{\sqrt{ 1 - \sin^2 2 \theta_{12} \, \sin^2 \alpha} 
- \meff/m_0}{1 + \sqrt{ 1 - \sin^2 2 \theta_{12} \, \sin^2 \alpha}}~.
\label{eq:QDlimit}
\ee
\end{itemize}
In the last two cases it is obvious that the limit is strongest 
for $\alpha = 0$, in the sense that more parameter space could be 
excluded for this value of $\alpha$. 
In Fig.~\ref{fig:NHlimits} we show the results of a detailed numerical 
analysis -- using the exact formulae -- of the possibility to 
obtain information on $|U_{e3}|^2$ 
with \onbb~in case of a normal mass ordering. 
Different experimental limits $m_{\rm exp}$ on the effective 
mass are assumed. 
The white area in Fig.~\ref{fig:NHlimits} 
is the one in which an upper or lower limit on $|U_{e3}|^2$ 
could be set, depending on the precise values of the oscillation parameters, 
neutrino masses, and Majorana phases. This area corresponds to 
the white 3$\sigma$-region in Fig.~\ref{fig:allowed}. 
It is separated by the dashed line 
from the green area (medium grey), which is marked with ``allowed'' and 
contains the allowed values of $m_1$ and $|U_{e3}|^2$. 
The green (medium grey) and white areas are the same as in the right plot of 
Fig.~\ref{fig:allowed}. The point is that for a limit on \meff, 
denoted here $m_{\rm exp}$, these areas become smaller because the 
yellow and red areas cut into them: The dark yellow 
(darkest grey) area marked by ``BF'' between the solid lines  
is the range of the lower limit 
on $|U_{e3}|^2$ if $\dms$, $\dma$, and $\theta_{12}$ are fixed to 
their best-fit values. If these parameters are allowed to vary within 
their $3\sigma$-ranges, then this area grows, and it is given by the 
light yellow (lightest grey) area between the dotted lines, marked with 
``$3\sigma$''.  
As mentioned above, the limit on $|U_{e3}|^2$ is a function of the 
Majorana phase $\alpha$ (cf.~Eq.~(\ref{eq:lowerlimit})). 
The lower limit is therefore not a sharp line, 
but an area because of its variation with $\alpha$. 
We indicated in the plot the extreme cases $\alpha = \pi/2$ and $\alpha=0$. 
As mentioned above, for the latter value more parameter space can be 
covered, as the respective curve (solid for the best-fit parameters, dotted 
for the $3\sigma$-ranges) is to the left of the line for  $\alpha=\pi/2$. 
The red (darkest) region is the one which is 
incompatible with the measured limit of \meff. If the limit 
on \meff~decreases, the excluded area increases of course. If the 
limit becomes very small, i.e., below 0.001 eV, then large part of 
the $m_1$-$|U_{e3}|^2$-plane can be excluded. In plots like the ones in 
Fig.~\ref{fig:meff} such small values of \meff~correspond to the 
``chimney'', in which the value of the effective mass sharply drops 
and only a certain range for $m_1$ is allowed.\\

\begin{figure}[tb]
\begin{tabular}[h]{lr}
\epsfig{file=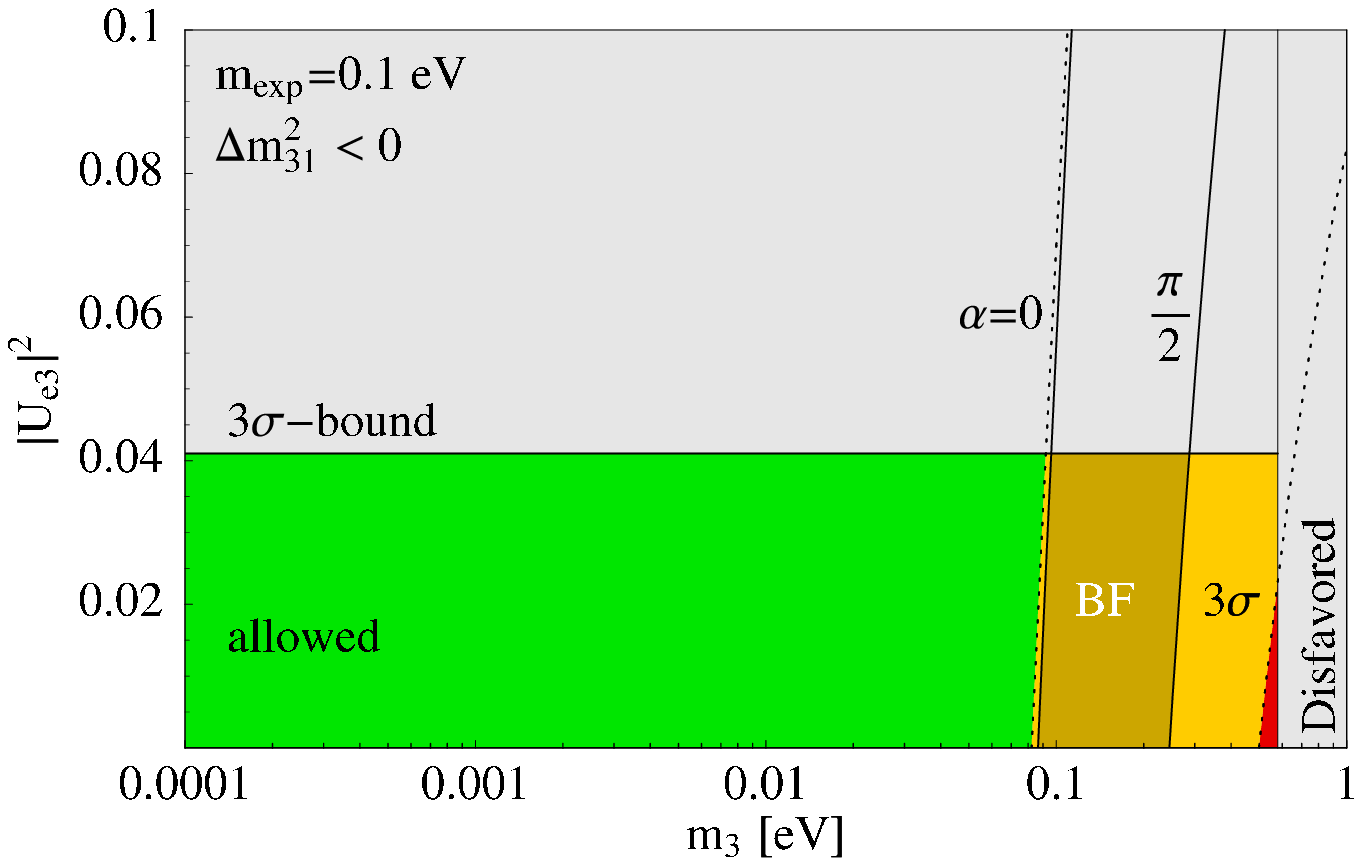,width=8cm}  &
\epsfig{file=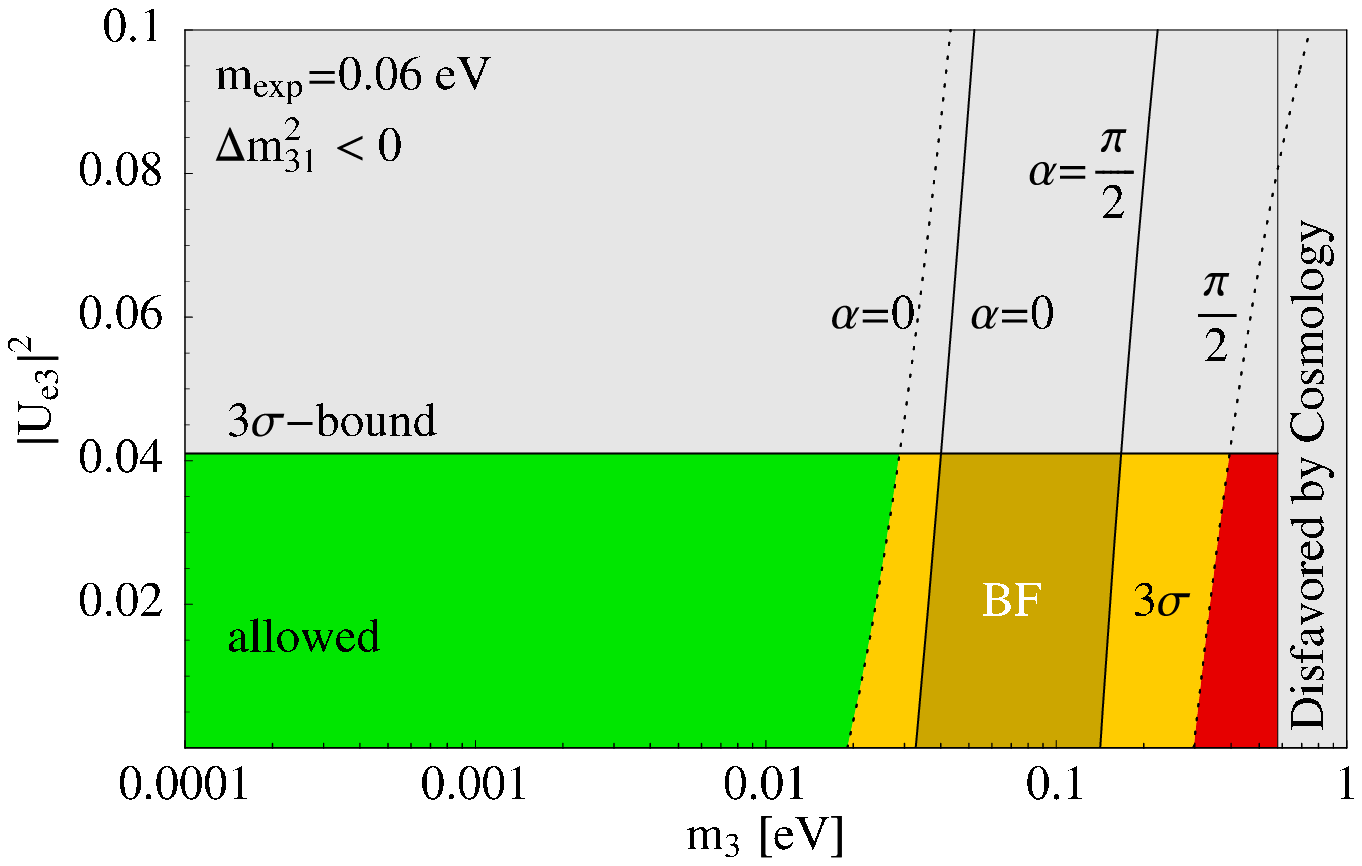,width=8cm} \\
\epsfig{file=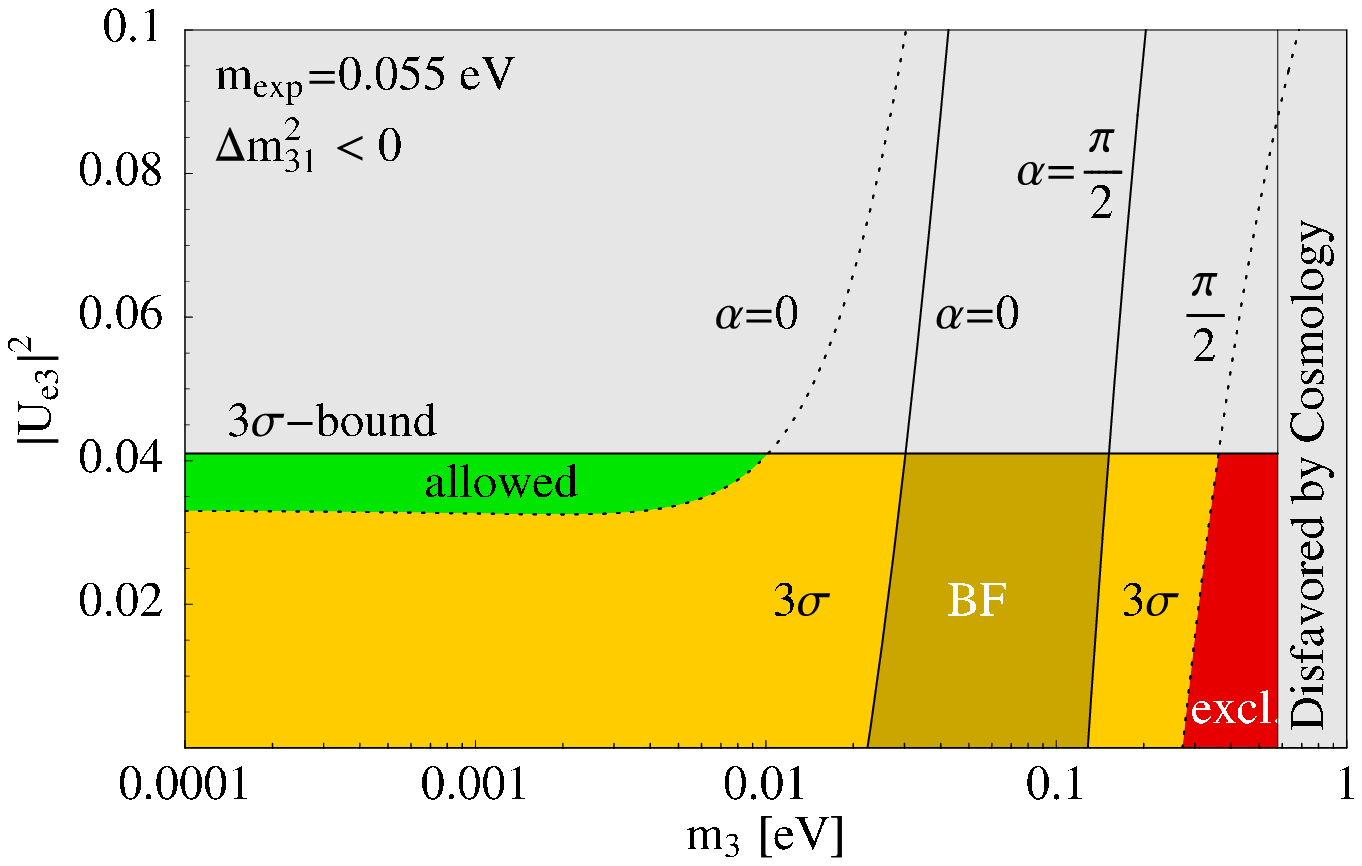,width=8cm} &
\epsfig{file=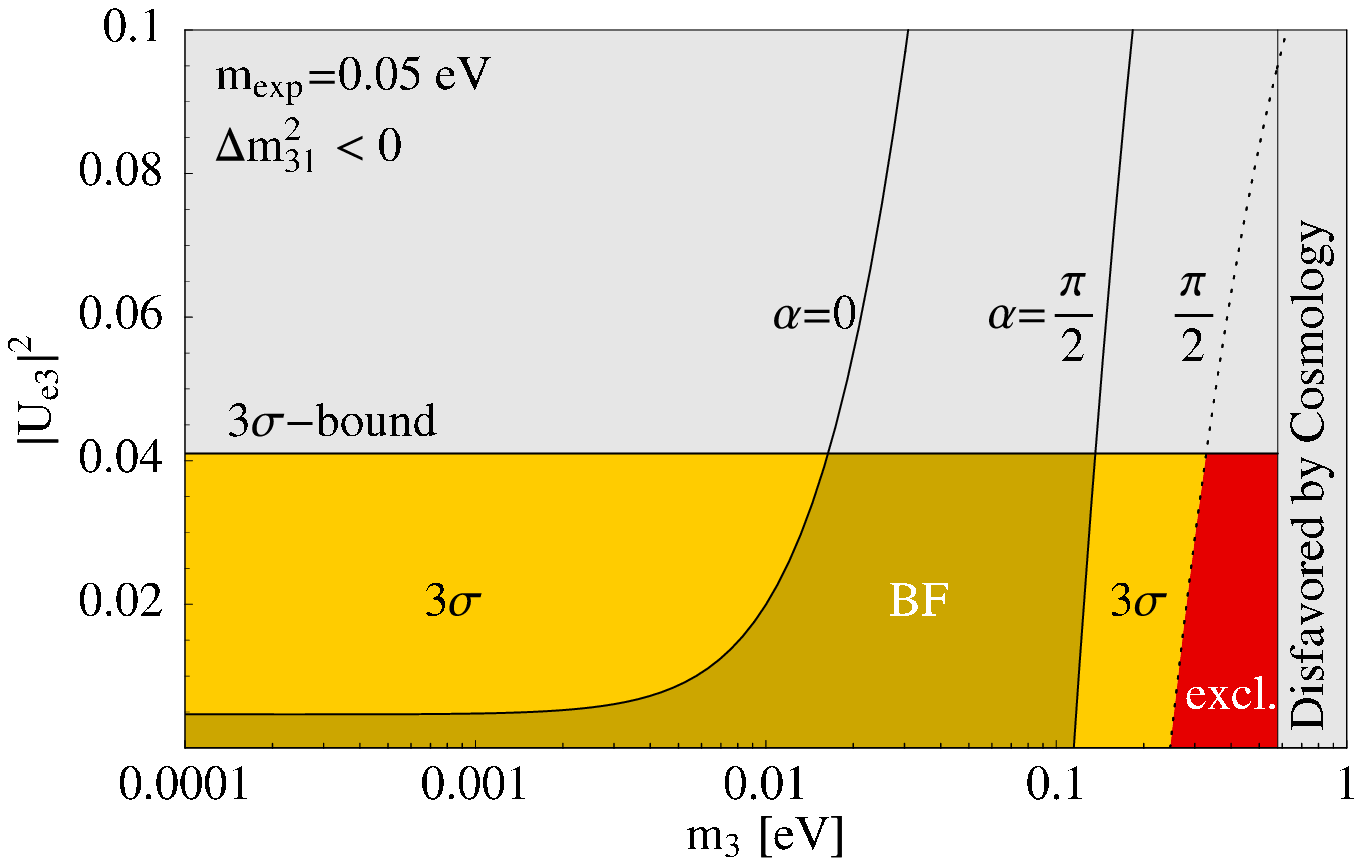,width=8cm} \\
\epsfig{file=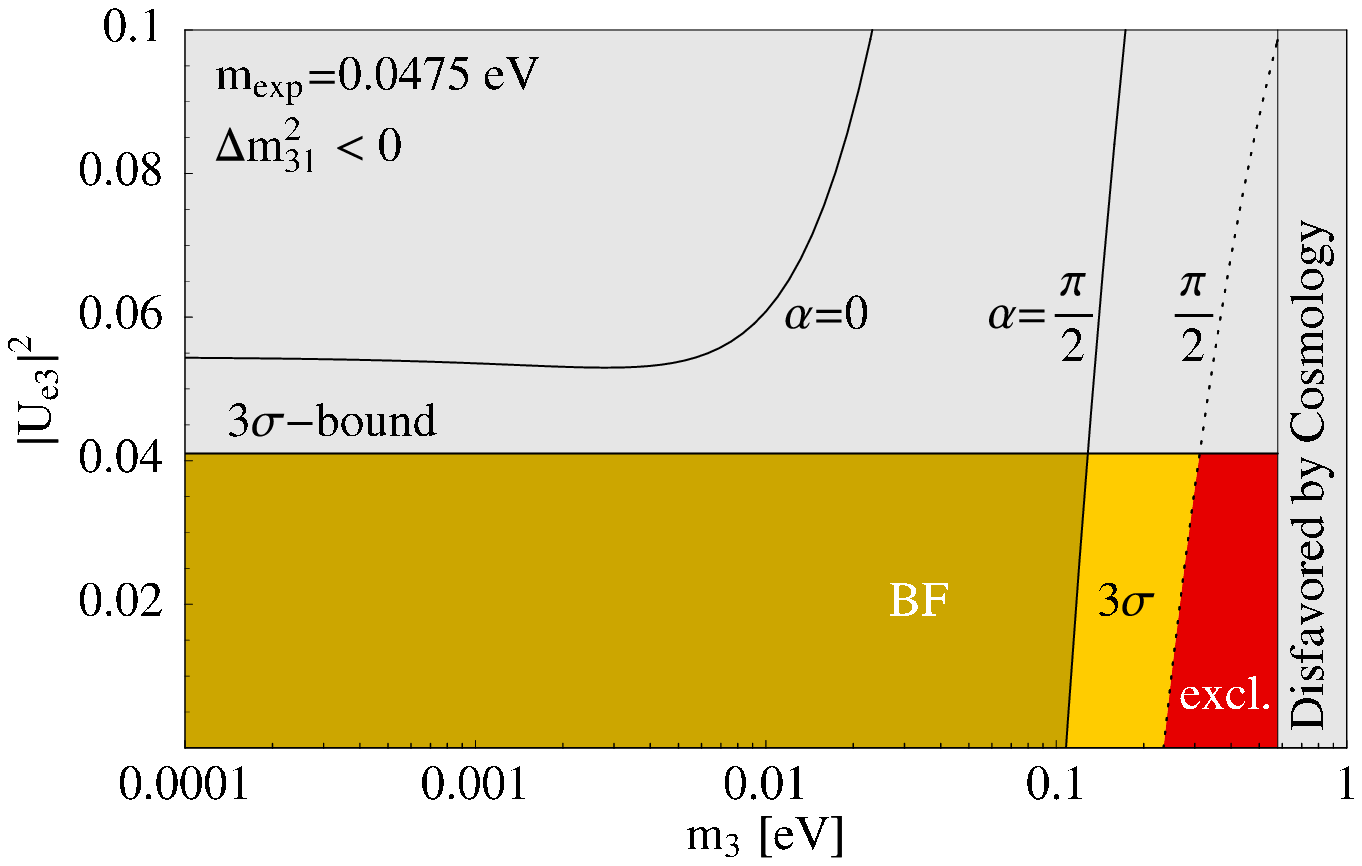,width=8cm} &
\epsfig{file=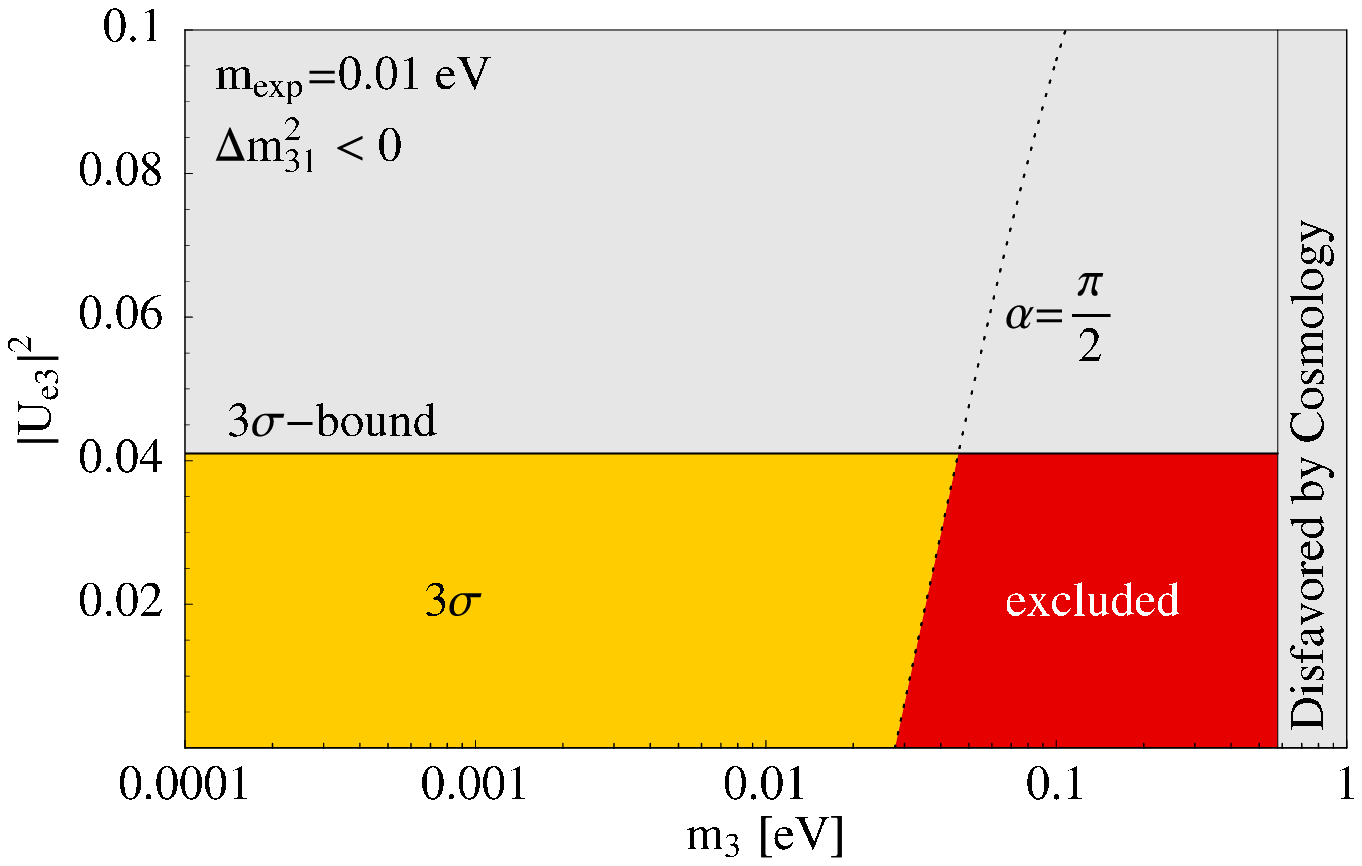,width=8cm}
\end{tabular}
\caption{\label{fig:IHlimits}Inverted neutrino mass ordering: 
Areas in the $m_3$-$|U_{e3}|^2$ parameter space in which information 
from \onbb~experiments could be gained. 
Different experimental bounds $m_{\rm exp}$ on the effective mass are shown, 
as well as the current $3\sigma$ limit on $|U_{e3}|^2$.  
The range of the lower limit on $|U_{e3}|^2$ is shown within solid
lines in dark yellow (darkest grey) for the best-fit values (``BF'') 
of the oscillation parameters. 
The light yellow (lightest grey) area within the dotted lines is the 
lower limit for the $3\sigma$-ranges. 
The left and right borders of these areas correspond to $\alpha = 0$ 
and $\alpha = \pi/2$, respectively. 
The red (darkest) areas are excluded, the green (medium grey) 
areas are allowed.}
\end{figure}

\thispagestyle{empty}

\begin{figure}[tb]
\begin{tabular}[h]{lr}
\vspace{-0.5cm}
\epsfig{file=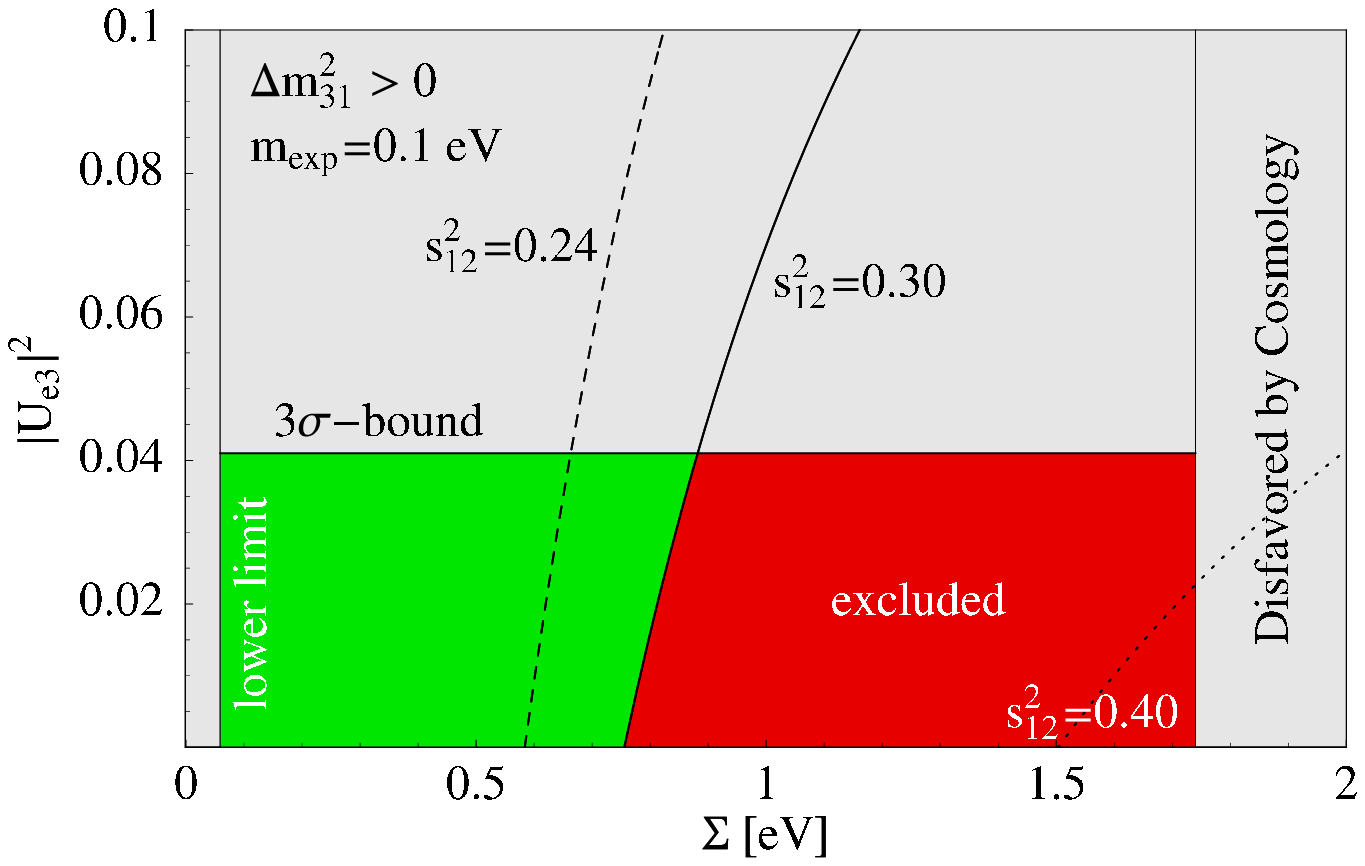,width=7.4cm}  &
\epsfig{file=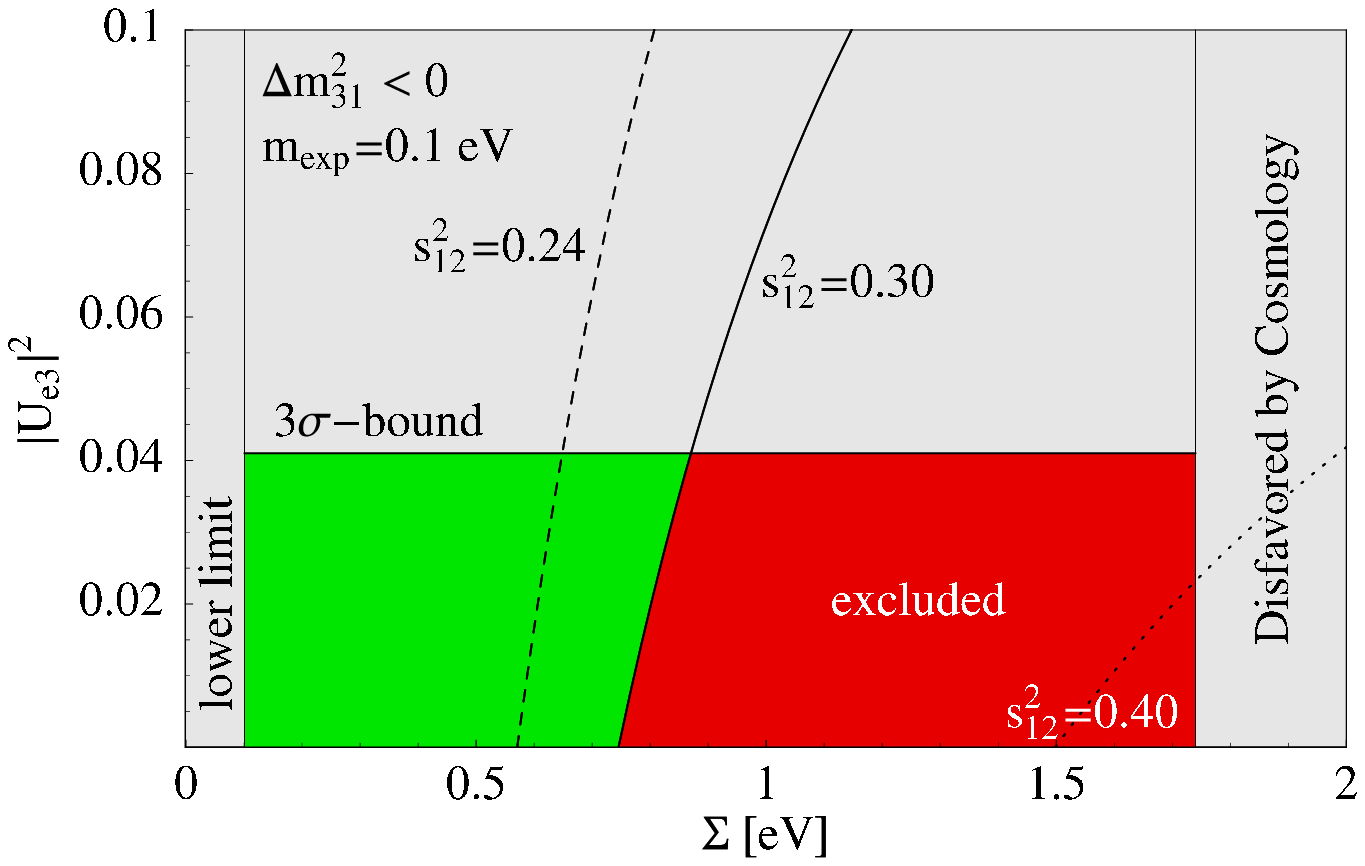,width=7.4cm} \\
\epsfig{file=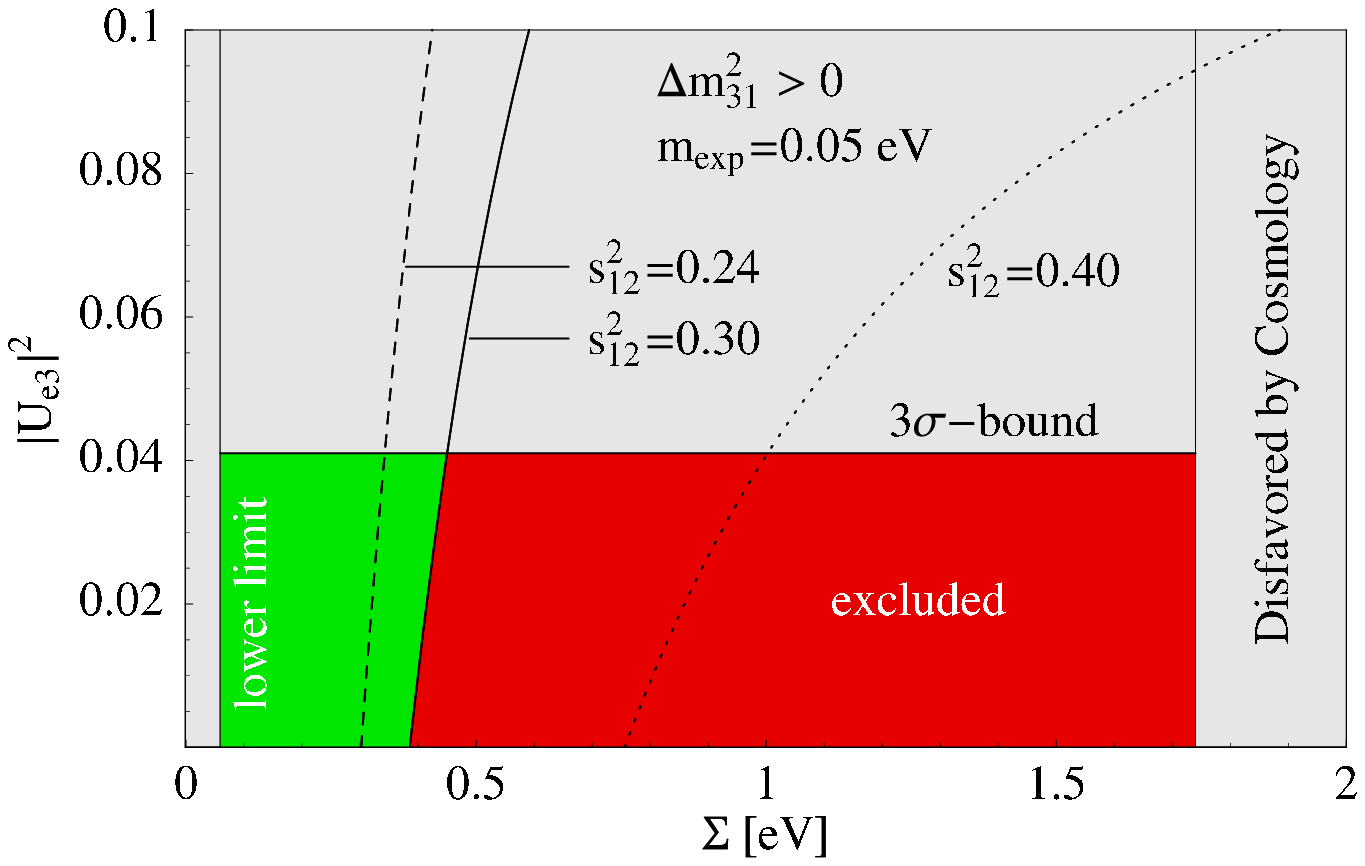,width=7.4cm} &
\epsfig{file=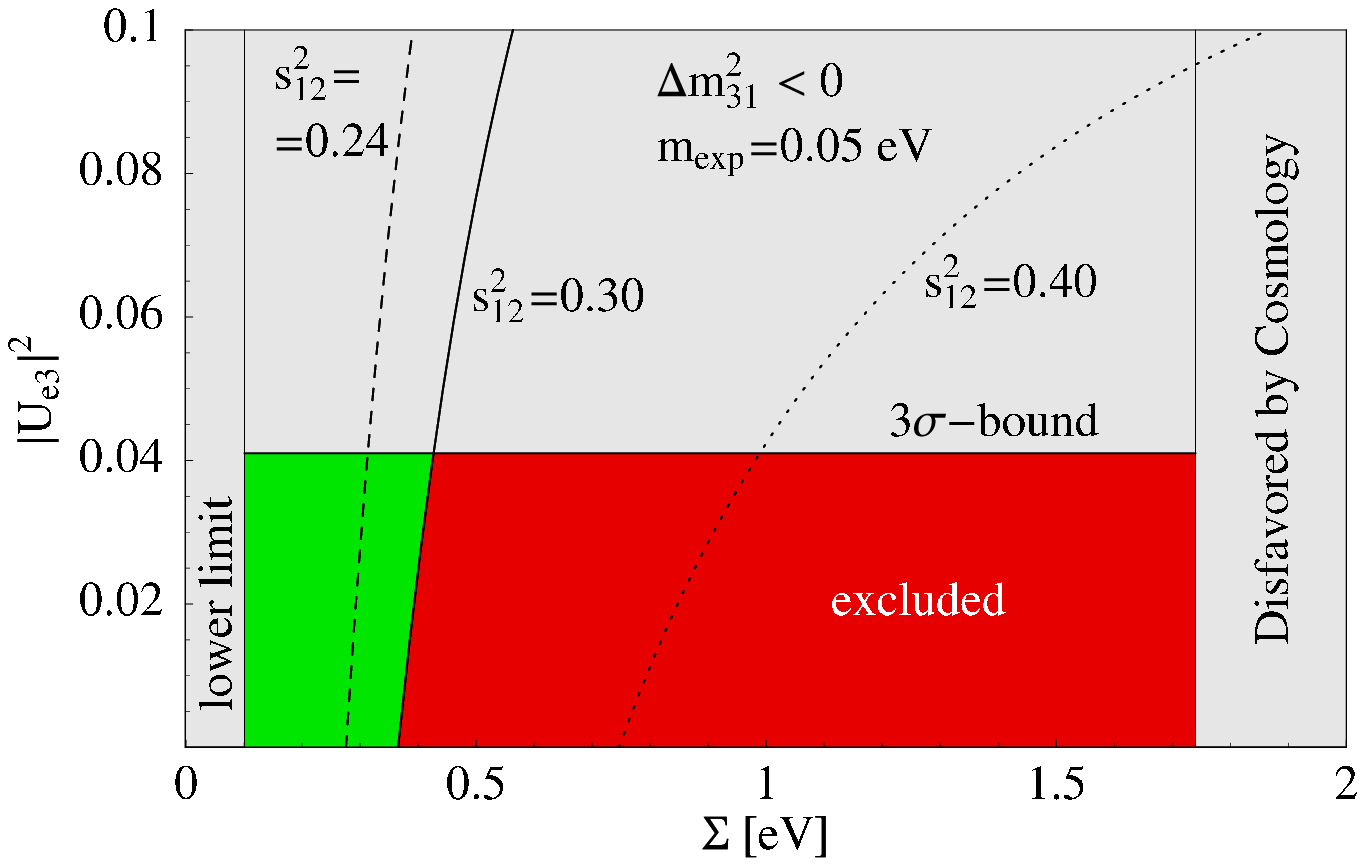,width=7.4cm} \\
\epsfig{file=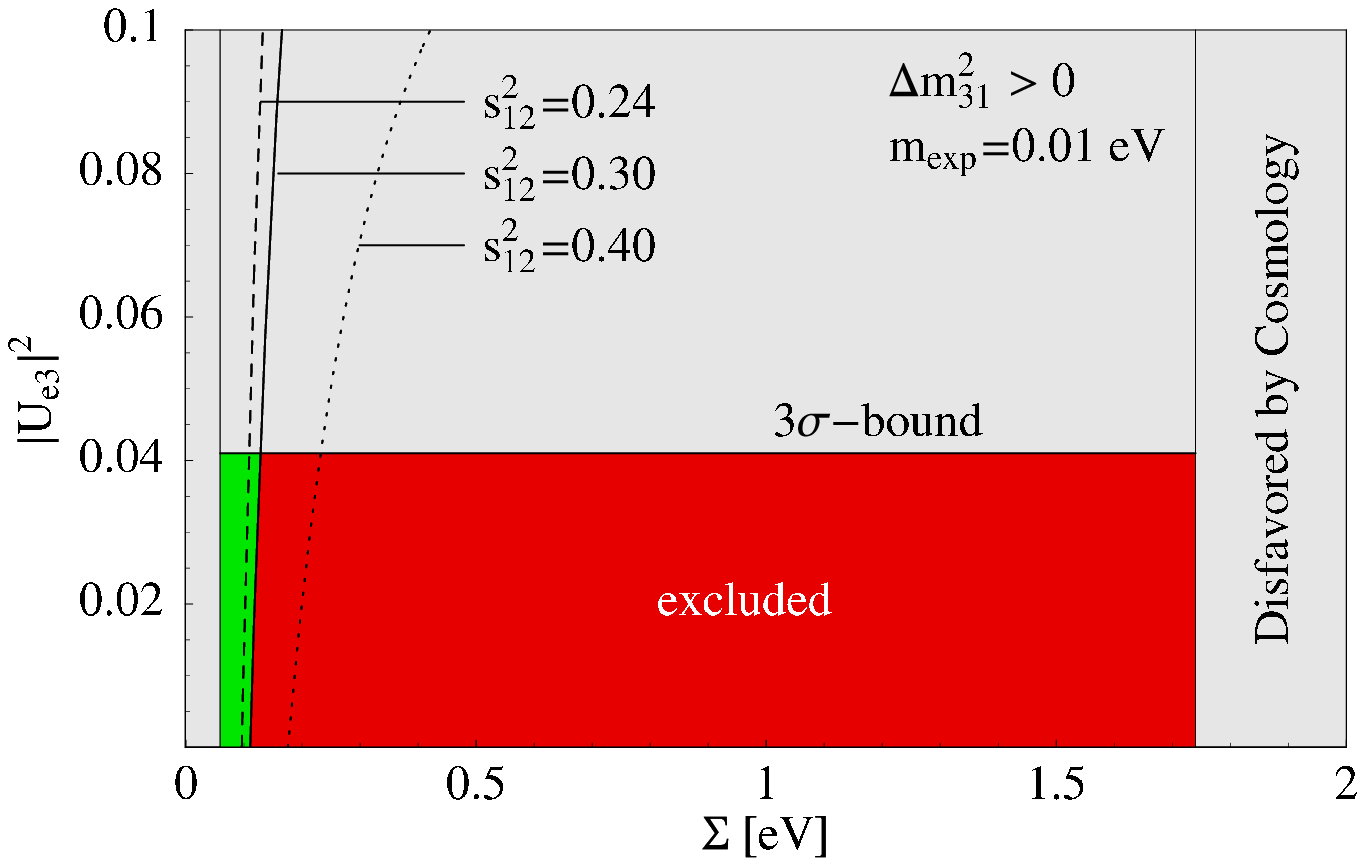,width=7.4cm} &
\epsfig{file=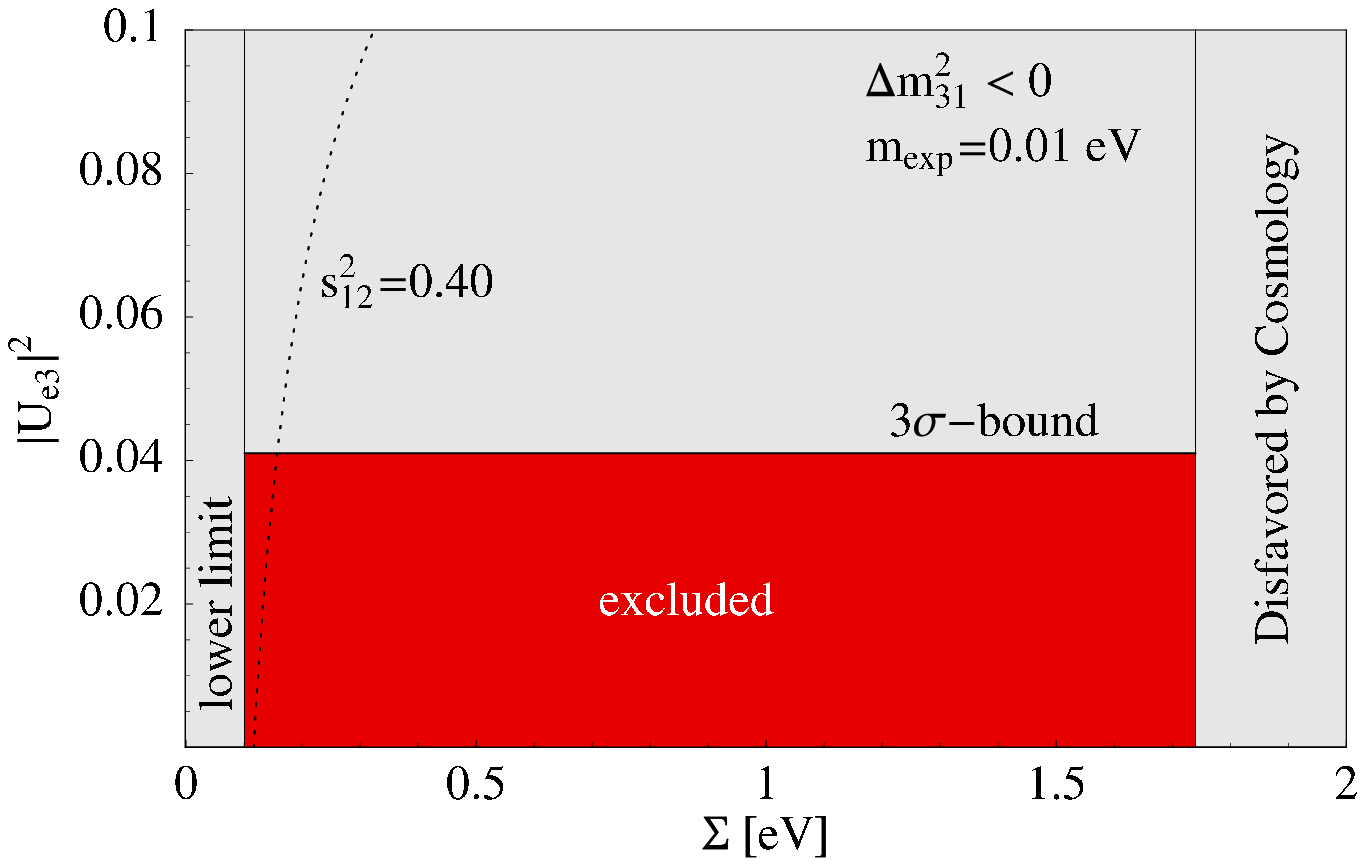,width=7.4cm} \\
\epsfig{file=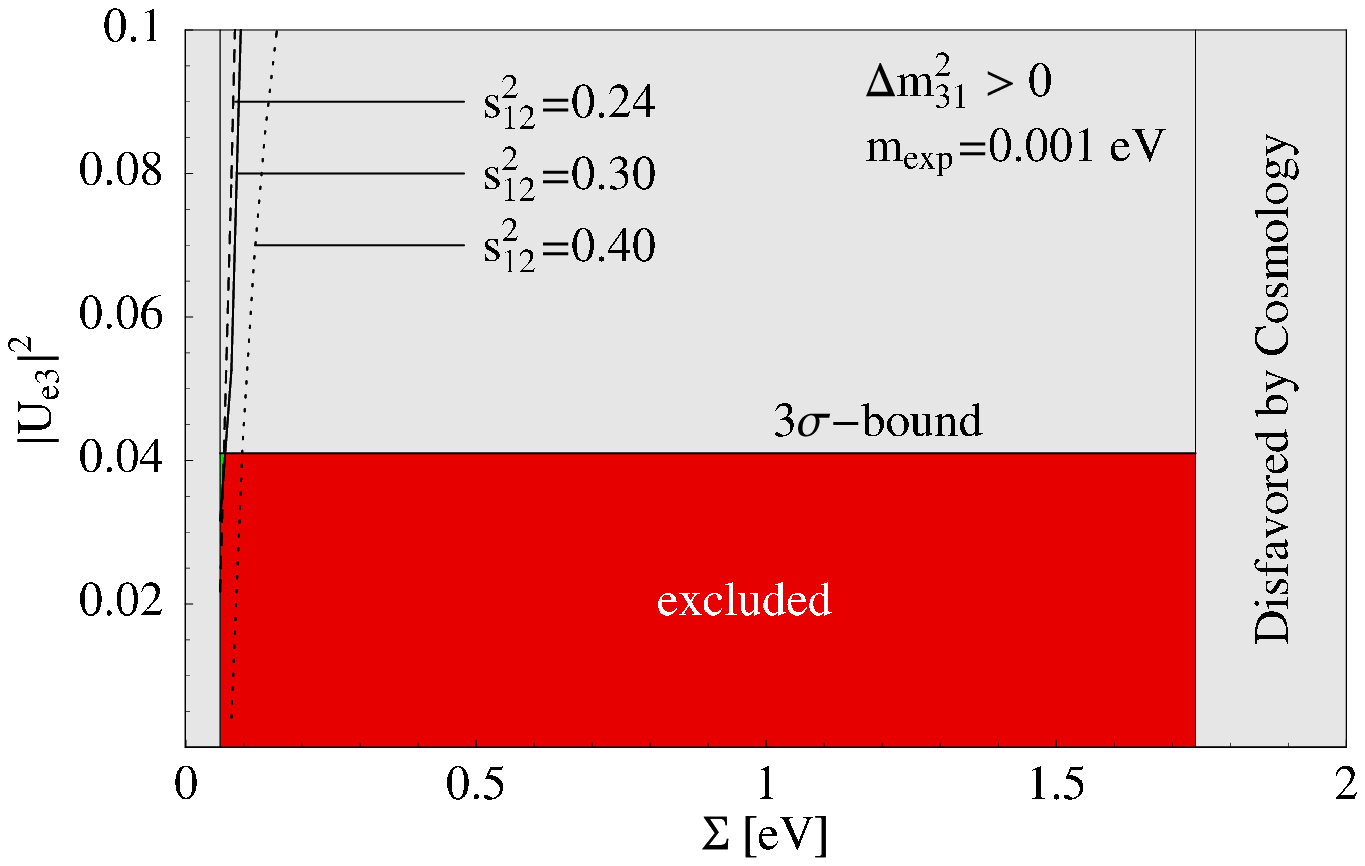,width=7.4cm} &
\end{tabular}
\caption{\label{fig:cosmo}
Excluded (red/dark grey) and allowed (green/medium grey) 
regions in the parameter space of $|U_{e3}|^2$ and 
$\Sigma = \sum m_i$ for a given limit $m_{\rm exp}$ on the effective mass and 
$\sin^2 \theta_{12} = 0.30$. 
The normal (inverted) mass ordering is shown in the left (right) panel. If 
$\sin^2 \theta_{12} = 0.24$, the regions would be separated by the dashed line, 
if $\sin^2 \theta_{12} = 0.40$, the regions would be separated 
by the dotted line. The current $3\sigma$ limit on $|U_{e3}|^2$, as well as 
the theoretical lower and the experimental upper limit on $\Sigma$ are 
also indicated.}
\end{figure}

Now we turn to the inverted mass ordering. Here the parameter 
space is defined by $m_3$ and $|U_{e3}|^2$. 
Fig.~\ref{fig:IHlimits} displays the result of our detailed analysis. 
The green (medium grey) area,  marked with ``allowed'', in Fig.~\ref{fig:IHlimits} 
is allowed by the respective limit $m_{\rm exp}$ on the effective mass. There 
is -- in contrast to the case of a normal hierarchy in 
Fig.~\ref{fig:NHlimits} -- no white area, because for the inverted ordering 
there is always a lower limit possible. 
The dark yellow (darkest grey) area marked by ``BF'' between the solid lines  
is the range of the lower limit on $|U_{e3}|^2$ if $\dms$, $\dma$, 
and $\theta_{12}$ are fixed to their best-fit values. 
If these parameters are allowed to vary within 
their $3\sigma$-ranges, then this area grows, and it is given by the 
light yellow (lightest grey) area between the dotted lines, marked with 
``$3\sigma$''.  
As the limit on $|U_{e3}|^2$ is a function of the 
Majorana phase $\alpha$, the lower limit is not a sharp line, 
but an area. We indicated in the plot the extreme 
cases $\alpha = \pi/2$ and $\alpha=0$, and again 
for the latter value more parameter space can be 
covered. Parameters lying in the red (darkest) region are 
incompatible with the measured limit of \meff. 
In general, the excluded area is larger than for the normal ordering 
shown in Fig.~\ref{fig:NHlimits}, as long as the limit on 
\meff\ is the same, e.g.\ for $m_{\rm exp}=0.01$~eV. 
However, for values of \meff\ smaller than around $0.008$~eV 
(cf.\ Fig.~\ref{fig:meff}), the inverted mass ordering is excluded anyway, 
and no information can be got for that case. 
For a limit on \meff~of 0.1 eV we are 
in the quasi-degenerate regime, and the corresponding plot looks 
identical to the one in the normal mass ordering in Fig.~\ref{fig:NHlimits}.

For a limit on \meff~of 0.01 eV there is no best-fit area anymore. The 
reason is that (neglecting $m_3 \, |U_{e3}|^2$) the effective mass 
is larger than 
$c_{13}^2 \, \sqrt{\dma} \, \cos 2 \theta_{12}$, which for the 
best-fit values of \dma~and $\theta_{12}$ is 
$0.02 \, c_{13}^2$ eV. With $c_{13}^2 \gs 0.96$ one cannot reach 0.01 eV. 
If the experimental limit becomes smaller than roughly 0.06 eV, one enters 
the horizontal band in Fig.~\ref{fig:meff}, in which the effective mass 
is basically independent of $m_3$. Decreasing the bound 
down to the minimal \meff~of 0.048 eV 
will therefore strongly constrain $|U_{e3}|^2$. 
These considerations are indeed confirmed by the plots.\\ 

In Fig.~\ref{fig:cosmo} we concentrate on the forbidden (red or dark grey) and 
allowed (green or medium grey) 
regions of $|U_{e3}|^2$ and of the sum $\Sigma = \sum m_i$ of neutrino masses.
Cosmological observations are sensitive to this observable, 
and current limits of order 1 eV are expected to be improved 
considerably \cite{steen}. Our quite conservative example bound 
is $\Sigma<1.74$~eV from Ref.~\cite{Tegmark:2003ud}. 
The current 3$\sigma$ limit on $|U_{e3}|^2$ is also indicated in the plots. 
We consider different limits $m_{\rm exp}$ on the effective mass and  
show both cases, normal and inverted mass ordering. 
There is a lower limit on $\Sigma = \sum m_i$ given by 
$\sqrt{\dma} + \sqrt{\dms}$ (normal ordering) and 
$2 \, \sqrt{\dma}$ (inverted ordering), also shown in the plot. 
The regions have been obtained using $\alpha=\frac{\pi}{2}$ and the 
best-fit value of $\sin^2 \theta_{12} = 0.30$. 
The choice $\alpha=\frac{\pi}{2}$ is the most 
conservative case in the sense that, for other values of $\alpha$, 
an even larger area could be excluded. In turn, the red (dark grey) 
region is forbidden independently of the exact value of the Majorana 
phase $\alpha$. We have also indicated how the areas change when 
$\sin^2 \theta_{12}$ is not the current best-fit value. For its upper and lower 
3$\sigma$ limits the allowed (red or dark grey) and forbidden 
(green or medium grey) areas would be separated by the 
dashed line ($\sin^2 \theta_{12} = 0.24$) and the dotted line 
($\sin^2 \theta_{12} = 0.40$).

Knowledge of the solar mixing angle is obviously needed to give stronger 
constraints. Note that the upper two plots in Fig.~\ref{fig:cosmo} 
have been drawn for an experimental 
limit on the effective mass of $m_{\rm exp}=0.01$~eV, which is expected 
to be achieved, e.g., by GERDA within 3 years of running \cite{Abt:2004yk}. 
Hence, after that period one could, with a better knowledge on $s_{12}^2$, 
exclude about half of the $|U_{e3}|^2$-$\Sigma$-parameter space (or even 
more) from a {\it non}-observation of \onbb. There is a strong dependence on 
$s_{12}^2$ for this value of $m_{\rm exp}$. 
For a better limit from 
double beta experiments, the exact value of the solar neutrino mixing 
angle would be less important to give constraints. The precision of 
$\theta_{12}$ is however quite important for limits on \meff\ in 
the regime of $0.1$~eV.  

The information obtained on $|U_{e3}|^2$ summarized in Figs.~\ref{fig:NHlimits}, \ref{fig:IHlimits}, and \ref{fig:cosmo} assumes that only light neutrino exchange is 
responsible for neutrino-less double beta decay. From those, 
Figure \ref{fig:cosmo} is better suited for consistency checks and 
looking for alternative mechanism of \obb. For instance, suppose that future 
data shows that $\sin^2 \theta_{12} = 0.30$ to a good precision 
(this is quite plausible as it is the best known mixing parameter). 
In addition, let us assume that a limit of 0.1 eV on the effective mass is established and that $\Sigma = 0.8$ eV is inferred from 
cosmology. If $|U_{e3}|^2 = 0.04$ is measured, then everything is consistent.  
In contrast, if a reactor neutrino experiment shows that $|U_{e3}|^2 = 0.01$,  
then we are in the excluded region. In this case -- assuming that neutrino 
oscillation physics is described by standard 3-flavor physics -- 
the following possibilities are present: 
(i) neutrinos are Dirac particles; (ii) cosmology has unexplained features 
mimicking effects of neutrino masses; 
(iii) there is a mechanism besides light neutrino exchange which 
contributes significantly 
to \obb~and whose amplitude interferes destructively with it. The first item 
could be checked by a KATRIN measurement corresponding to a neutrino mass 
of $\simeq \Sigma/3 \simeq 0.27$, which is above the claimed sensitivity of 
0.2 eV \cite{KATRIN}. The last item might be indirectly checked if R-parity 
violating SUSY is contributing to \obb~\cite{SUSY}. In this case, the 
relevant parameters such as sparticle masses and their coupling constants 
with particles may be inferred from collider experiments. 
Many more examples in this spirit can be read off from Fig.~\ref{fig:cosmo}.\\

It is now the right time to mention the uncertainty stemming from 
the Nuclear Matrix Elements (NMEs), which will be the 
dominating error in what regards the extraction of 
physics parameters from \meff. We can write the decay width of 
\onbb~as $\Gamma = x \, \meff^2$, where $x = G(E_0, Z) \, |{\cal M}|^2$ 
contains the phase space factor $G(E_0, Z)$ -- which is a calculable 
function of the nucleus -- and the nuclear 
matrix element $|{\cal M}|^2$. If $\sqrt{x}$ 
goes from $y$ to $\zeta \, y$, where $\zeta$ is the NME uncertainty, then 
the uncertainty on \meff~is \cite{Choubey:2005rq,PPR0}
\be \label{eq:sigma_meff}
\Delta(\meff) \equiv \frac{\meff^{\rm max} - \meff^{\rm min }}
{\meff^{\rm max} + \meff^{\rm min}} = \frac{\zeta - 1}{\zeta + 1}~.
\ee
Of course, in what regards the limit on $|U_{e3}|^2$, the NME uncertainty 
can simply be taken into account by modifying the experimental limit 
on \meff~accordingly, 
which enables us to read off the influence of such uncertainties directly 
from Figs.~\ref{fig:NHlimits} and \ref{fig:IHlimits}. Recall that the bounds on 
$|U_{e3}|^2$ are very sensitive 
to the limit of \meff~in case of an inverted hierarchy. 
For instance, in Fig.~\ref{fig:IHlimits} the change between the limits 
$\meff \le 0.06$ eV and $\meff \le 0.05$ eV is -- as discussed above -- 
quite dramatic. On the other hand, between the values 
$\meff \le 0.1$ eV and $\meff \le 0.06$ eV there is not much difference. 
Nevertheless, this indicates that the 
possible impact of the NME uncertainty 
in what regards information on $|U_{e3}|^2$ depends 
crucially on the value of the effective mass. We can estimate that in case the inverted mass hierarchy regime 
is touched by experiments ($\meff$ smaller than 0.05 eV), NME uncertainties 
smaller than 20\% are necessary for robust limits and 
exclusion areas. For smaller limits on $\meff$, 
uncertainties up to a factor 2 are still allowed, as they render the 
excluded areas rather robust.\\

\begin{figure}[tb]
\begin{tabular}[h]{lr}
\epsfig{file=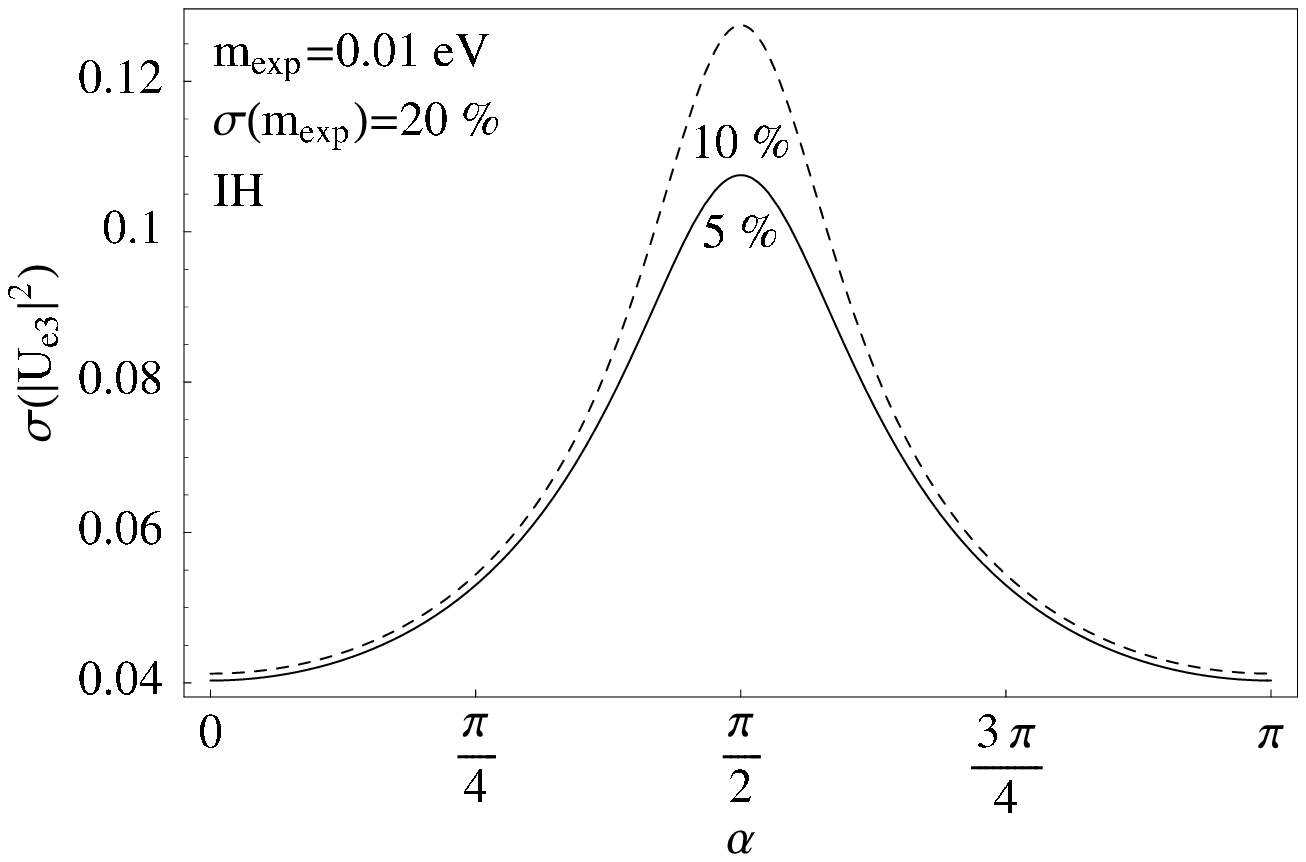,width=8cm} &
\epsfig{file=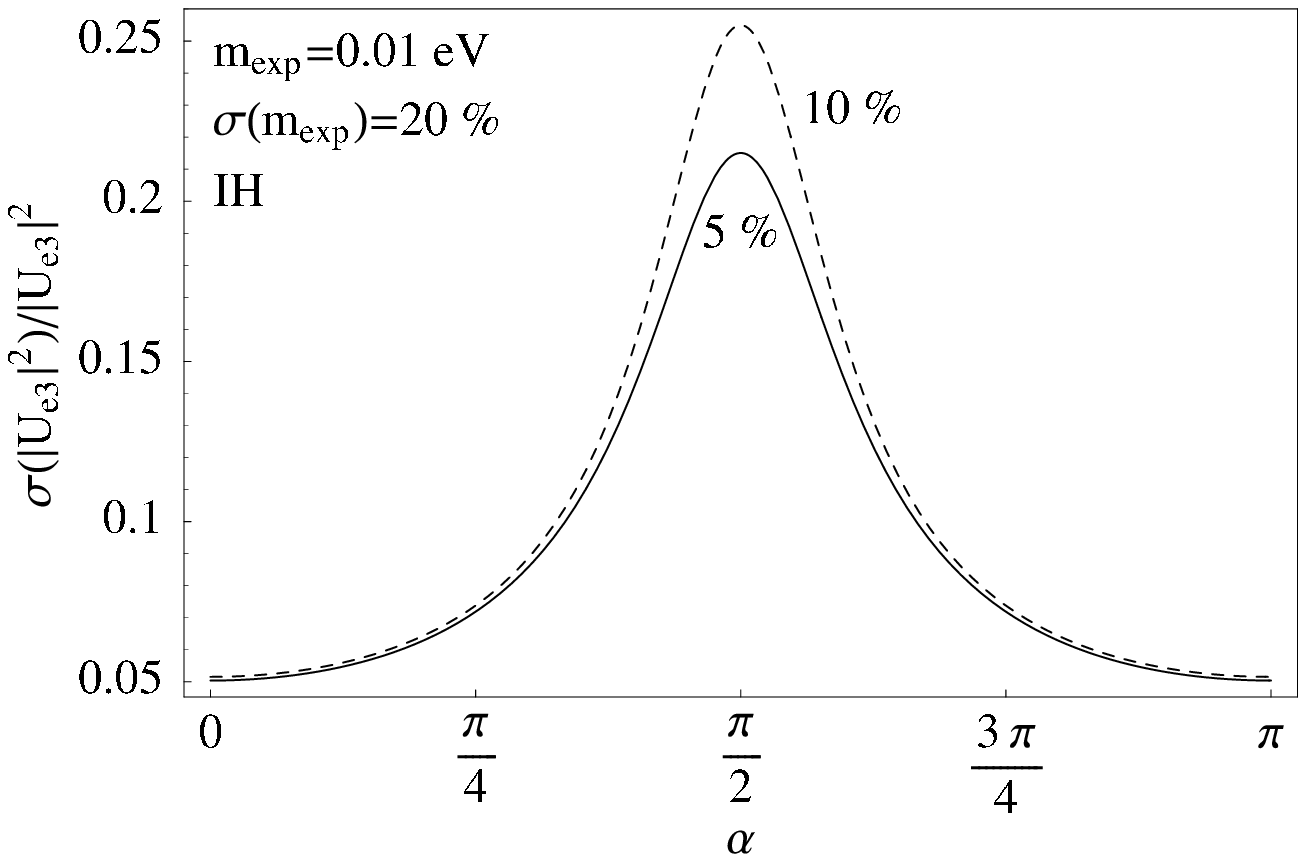,width=8cm}
\end{tabular}
\caption{\label{fig:ue3_sig1}
The absolute and relative error on $|U_{e3}|^2$ in case of an inverted 
hierarchy for negligible $m_3 \, |U_{e3}|^2$.\ \dma~and $\sin^2 \theta_{12}$ 
are both given a relative error of 5 and 10~\%, respectively. The strong 
dependence on $\alpha$ is clearly visible from the plots.}
\end{figure}

Much more crucial will the uncertainty be if one wants to {\it measure} 
$|U_{e3}|^2$. As an example, consider the inverted hierarchy in the 
limit of vanishing $m_3 \, |U_{e3}|^2$. As can be seen from 
Fig.~\ref{fig:meff}, this approximation is valid for $m_3 \ls 0.01$ eV. 
In this case the formula 
\be
|U_{e3}|^2 = 1 - \frac{\meff}{\sqrt{\dma} \, 
\sqrt{1 - \sin^2 2 \theta_{12} \, \sin^2 \alpha}}~
\label{eq:Ue3meas}
\ee
can easily be obtained. Simple error propagation yields 
\bea \label{eq:sig_ue3} \D 
\sigma(|U_{e3}|^2) = 
\frac{\meff}{2 \, \sqrt{\dma} 
\, (1 - \sin^2 2 \theta_{12} \, s_\alpha^2)^{3/2}} \, 
\left[
\left(
1 - \sin^2 2 \theta_{12} \, s_\alpha^2\right)^2 
\left(\frac{\sigma(\dma)}{\dma} \right)^2 +
\right. \\[0.2cm] \D 
\left. 
+ 
4 \left(4 \cos^2 2 \theta_{12} \, s_{12}^4 \, 
s_\alpha^4  \, 
\left(\frac{\sigma(s_{12}^2)}{s_{12}^2} \right)^2
+  
\left(1 - \sin^2 2 \theta_{12} \, s_\alpha^2
\right)^2 \, \left(\frac{\sigma(\meff)}{\meff} \right)^2 
\right)\right]^{1/2}~.
\eea
In Fig.~\ref{fig:ue3_sig1} we plot $\sigma(|U_{e3}|^2)$ and 
$\sigma(|U_{e3}|^2)/|U_{e3}|^2$ while 
taking relative errors of 5 and 10~\% for \dma~and 
$\sin^2 \theta_{12}$, respectively \cite{fut_osc}. In what regards the 
effective mass, we assume an experimental 
error $\sigma(\meff)/\meff$ of 20 \%. The error shows a strong dependence 
on the Majorana phase $\alpha$. Interestingly, especially the value of 
$\alpha=\frac{\pi}{2}$, which allowed to set the realistic lower limit 
on $|U_{e3}|^2$ for a non-observation of \onbb, is the most conservative case 
for a $|U_{e3}|^2$-measurement. The relative error would be nearly 25~\% in 
that case, but can also go down to approximately 5~\% for $\alpha=0$, which 
would then be comparable to the current knowledge of the oscillation parameters. 
However, as the absolute error on $|U_{e3}|^2$ is larger than the currently 
allowed 3$\sigma$-bound, we conclude that a measurement of this parameter 
via $0\nu\beta\beta$ is very difficult, simply due to lack of 
knowledge on the Majorana phase $\alpha$. 

\begin{figure}[tb]
\begin{tabular}[h]{lr}
\epsfig{file=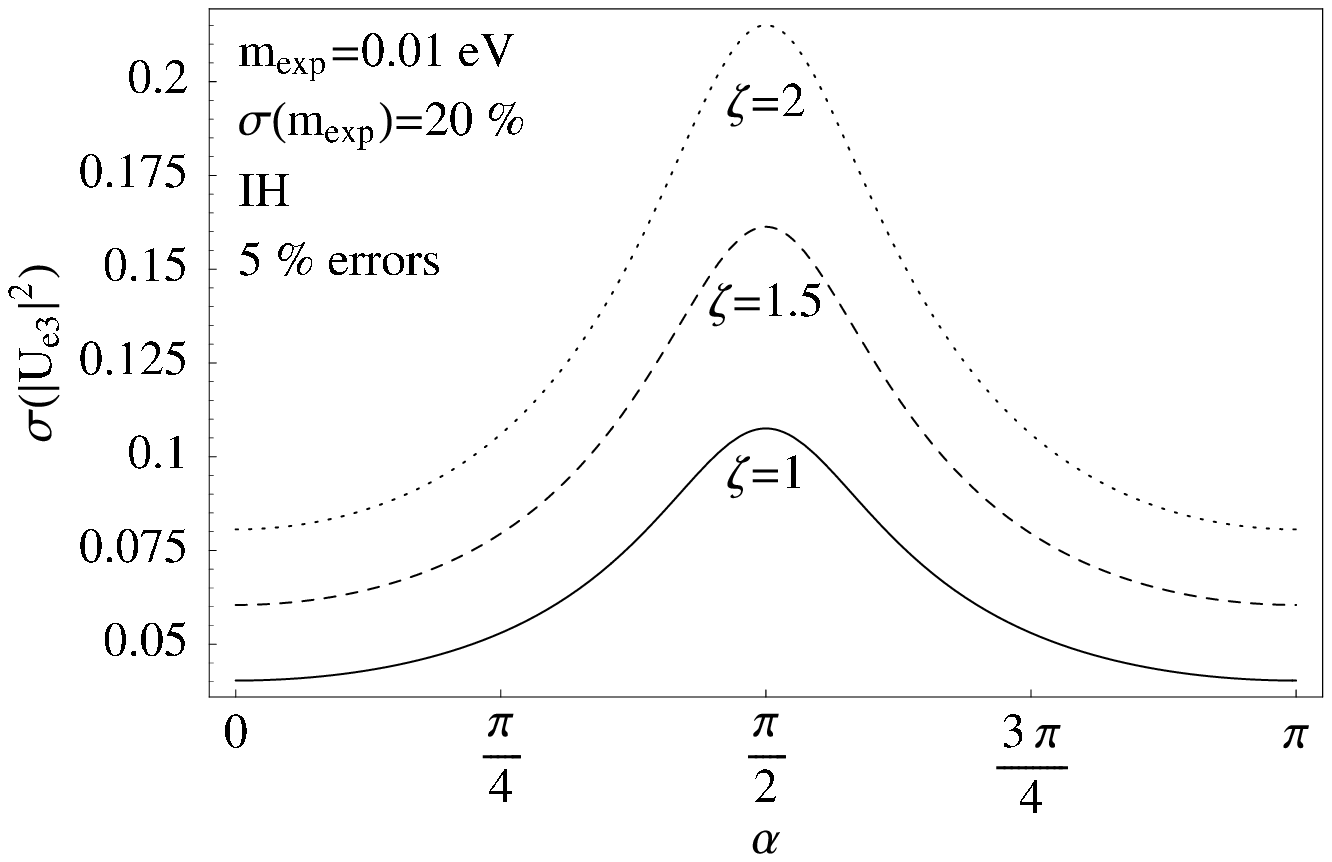,width=8cm} &
\epsfig{file=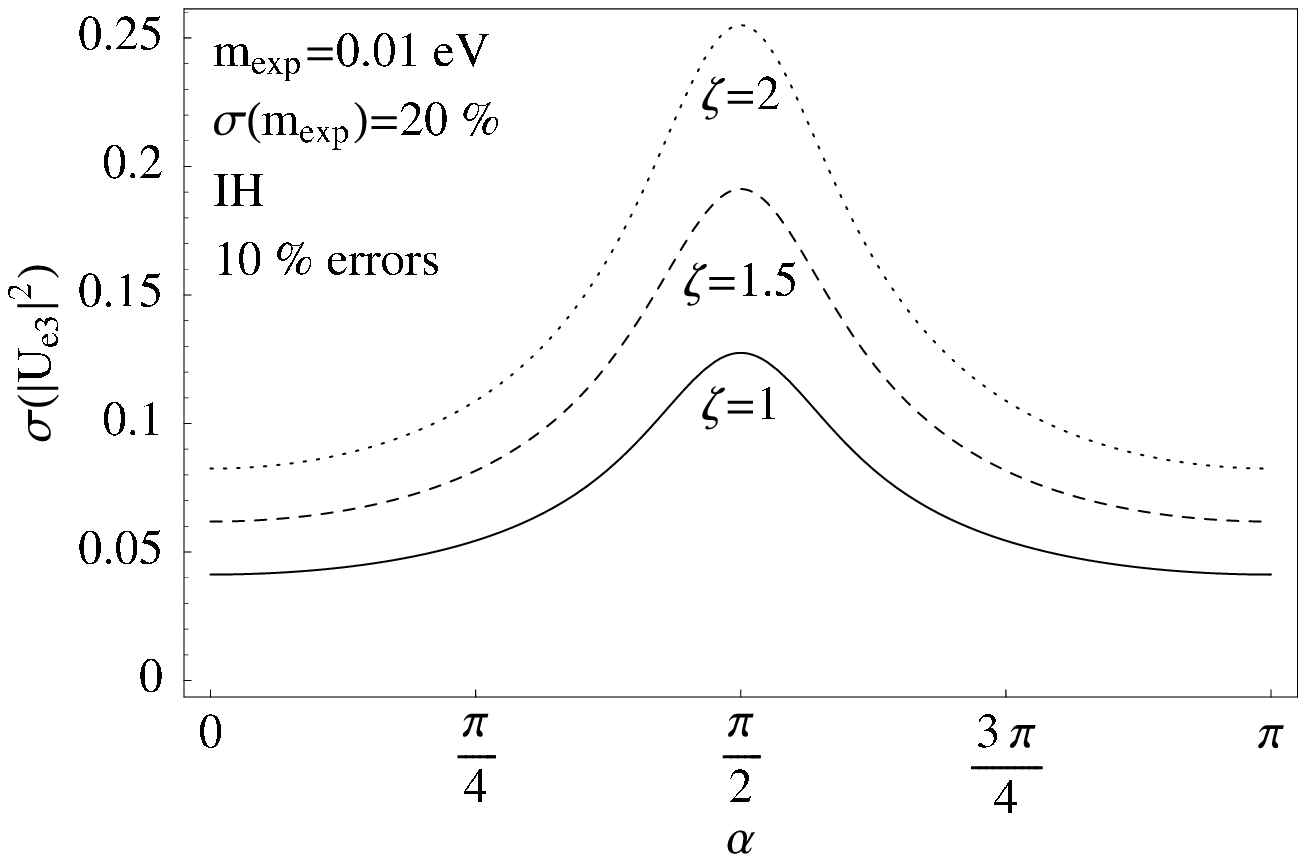,width=8cm}
\end{tabular}
\caption{\label{fig:zeta}
The influence of the nuclear matrix element uncertainty $\zeta$ on the 
 error of $|U_{e3}|^2$ in case of an inverted hierarchy. The errors on 
\dma~and $\sin^2 \theta_{12}$ are again 5 and 10 \%, respectively. 
$\sigma(|U_{e3}|^2)$ grows linearly with $\zeta$.}
\end{figure}

The NME uncertainty $\zeta$ can be 
taken into account by using \cite{PPR0} 
$\meff = \zeta \, \meff_{\rm min}^{\rm exp}$, 
where $\meff_{\rm min}^{\rm exp}$ is obtained from the measurement 
of \obb~when the largest NME is used. Then, one simply gets a factor of 
$\zeta$ in front of Eq.~(\ref{eq:sig_ue3}), which can maybe double the error, 
but does not change the order of magnitude. This is illustrated in 
Fig.~\ref{fig:zeta}, where we show $\sigma(|U_{e3}|^2)$ as a function of 
$\alpha$ and $\zeta$ for two examples of the errors on the oscillation 
parameters. However, for advantageous values of $\alpha$, one can get a 
reasonable measurement of $\sigma(|U_{e3}|^2)$, but therefore, external 
information on the Majorana phase would be needed to draw reasonable 
conclusions. Hence, measuring $|U_{e3}|$ from \onbb\ will be much more 
difficult than simply excluding part of the parameter space from a 
non-observation. In particular, $|U_{e3}|^2$ should lie very close 
to its currently allowed 3$\sigma$-limit.

%

\section{\label{sec:Conclusions}Conclusions}

Assuming that neutrinos are Majorana particles, we have 
performed a detailed analysis of what information on the 
lepton mixing matrix element $|U_{e3}|^2$ can be gained from a (non-) observation 
of \onbb\ in future experiments. After reviewing the current knowledge on 
the effective mass \meff, we have shown how one can get a {\it lower} limit 
on the mixing matrix element $|U_{e3}|^2$, which in particular depends on 
the neutrino mass and the Majorana phase $\alpha$. However, even after 
including the variation by the phase as well as by the uncertainties in 
the oscillation parameters, there still remains a sizable excluded area 
in the parameter space. We have presented this for the two cases of normal 
and inverted mass ordering and we have also shown the impact on the sum of 
neutrino masses, as measured in cosmological observations. Especially for 
the last point, a larger area of the parameter space could be excluded in 
the nearer future. The precision of $\theta_{12}$ is quite important for 
limits on \meff\ in the regime of $0.1$~eV, but much less important for 
smaller limits. We have also commented on the impact of errors in the 
Nuclear Matrix Element calculations. Depending on the specific case, 
this impact can be strong or weak. We have closed with showing that 
the ``inverse'', namely a measurement of the mixing matrix element 
$|U_{e3}|^2$ by an observation of \onbb, is much more difficult since 
the errors especially due to the variation of the Majorana phase $\alpha$ 
are quite large.

\section*{\label{sec:Ack}Acknowledgments} 
This work has been supported by the 
DFG-Sonderforschungsbereich Transregio 27 ``Neutrinos and 
beyond -- Weakly interacting particles in Physics, 
Astrophysics and Cosmology'' and under project 
number RO--2516/3-2 (W.R.), as well as by the 
EU program ILIAS N6 ENTApP WP1.


%

\end{document}